\documentclass[11pt,a4paper]{article}
\usepackage{jheppub}
\usepackage[utf8]{inputenc}
\usepackage{amsmath}
\usepackage{amssymb}
\usepackage{bbm}
\usepackage{bm}
\usepackage{bbold}
\usepackage{braket}

\def\be{\begin{equation}}
\def\ee{\end{equation}}
\def\bes{\begin{equation*}}
\def\ees{\end{equation*}}
\def\bea{\begin{eqnarray}}
\def\eea{\end{eqnarray}}
\def\bal{\begin{align}}
\def\eal{\end{align}}

\def\f{\frac}
\def\mc{\mathcal}
\def\mf{\mathfrak}
\def\mbb{\mathbb}
\def\mbs{\mathbbmss}

\def\v[#1]{\boldsymbol{#1}}
\def\w[#1]{\widehat{#1}}
\def\vs[#1,#2]{\boldsymbol{{#1}_{#2}}}
\def\mes[#1]{d^{3}{#1}}
\def\del{\partial}
\def\<{\langle}
\def\>{\rangle}
\def\vecs[#1,#2]{\boldsymbol{{#1}_{#2}}}

\newcommand{\half}{\frac{1}{2}}

\newcommand{\eq}[1]{eq.\ \eqref{#1}}
\newcommand{\Eq}[1]{Eq.\ \eqref{#1}}

\def\a{\alpha}
\def\b{\beta}
\def\d{\delta}
\def\D{\Delta}
\def\e{\epsilon}
\def\ve{\varepsilon}

\def\l{\lambda}
\def\L{\Lambda}
\def\m{\mu}
\def\n{\nu}
\def\N{\nabla}
\def\o{\omega}
\def\O{\Omega}

\def\vp{\varphi}
\def\r{\rho}
\def\s{\sigma}
\def\vsi{\varsigma}

\def\th{\theta}

\def\z{\zeta}

\newcommand{\tD}{\tilde{\D}}

\newcommand{\te}{\tilde{\epsilon}}
\newcommand{\tr}{\tilde{\r}}

\def\ET{{\cal E}}
\def\PT{{\cal P}}
\def\UT{{\cal U}}
\def\VT{{\cal V}}
\def\QT{{\cal Q}}
\def\RT{{\cal R}}
\def\TT{{\cal T}}
\def\NJ{{\cal N}}
\def\SJ{{\cal S}}
\def\JJ{{\cal J}}
\def\MX{{\cal M}}
\def\ZX{{\cal Z}}
\def\XX{{\cal X}}

\newcommand{\lr}[1]{\left(#1\right)}
\newcommand{\pder}[2]{\frac{\partial #1}{\partial #2}}

\def\kv{\mathbf{k}}
\def\xv{\mathbf{x}}

\title{Causal and Stable Superfluid Hydrodynamics}

\author[a]{Raphael E. Hoult}
\author[b,c,d]{and Ashish Shukla}

\affiliation[a]{Department of Physics \& Astronomy, University of Victoria, PO Box 1700 STN CSC, Victoria, BC V8W 2Y2, Canada}
\affiliation[b]{Theory Division, Saha Institute of Nuclear Physics, 1/AF Bidhan Nagar, Kolkata 700064, India.}
\affiliation[c]{Homi Bhabha National Institute, Anushakti Nagar, Mumbai 400094,
India.}
\affiliation[d]{CPHT, CNRS, \'{E}cole polytechnique, Institut Polytechnique de Paris, 91120 Palaiseau, France.}

\emailAdd{rhoult@uvic.ca}
\emailAdd{ashish.shukla@saha.ac.in}

\abstract{
We investigate the linearized stability and causality properties of relativistic viscous superfluid hydrodynamics. The Landau-Lifshitz-Clark-Putterman formulation for the theory of relativistic viscous superfluids suffers from the same instability and acausality issues as the relativistic Navier-Stokes equation for normal fluids when written in the formulations of Eckart or Landau and Lifshitz. We show that conditions to ensure stability and causality can be satisfied with judicious redefinitions of the hydrodynamic variables. The conditions we obtain hold at non-zero superfluid velocity as well.}

\preprint{CPHT-RR067.092024}

\begin{document}

\maketitle
\flushbottom

\section{Introduction}
\label{sec:Intro}
Hydrodynamics has been remarkably successful as an effective macroscopic description for physical systems near thermal equilibrium \cite{Landau-Lifshitz, Kovtun:2012rj, Jeon:2015dfa, Romatschke:2017ejr,Rezzolla-Zanotti}. This is so because on scales much larger than the microscopic scales associated to the system, such as the mean free path and mean free time, the relaxation dynamics of the out of equilibrium state is dominated by the transport of conserved quantities associated with the symmetries of the system. The hydrodynamic equations are consequently just the conservation equations for these quantities. For instance, the invariance of a system under spacetime translations leads to the conservation of energy and momentum, and similar dynamical equations follow for any other symmetries the system might have. The system of hydrodynamic equations only becomes closed once constitutive relations for the currents associated with conserved charges are provided in terms of the hydrodynamic variables, such as the temperature $T$ and the fluid velocity $u^\m$. Since the state is near equilibrium, one makes the key assumption that the hydrodynamic variables vary slowly in space and time. The constitutive relations can therefore be written in a derivative expansion, with each successive term carrying a higher number of derivatives acting on the hydrodynamic variables, and consequently getting smaller. The derivative expansion can in principle be continued to any desired order.

Upon truncation of the derivative expansion at a finite order, the formulation of hydrodynamics boils down to a set of coupled partial differential equations (PDEs) for the evolution of the hydrodynamic variables. In evaluating the dynamics of such a system, there is always a concern as to whether the resulting system of equations admits unphysical solutions which are unstable and/or acausal. For instance, when working in the standard Landau-Lifshitz~\cite{Landau-Lifshitz} or Eckart~\cite{Eckart-paper,weinberg:1972} formulations of relativistic first order hydrodynamics (i.e.~the relativistic Navier-Stokes equations), it is known that the equations admit unstable and acausal solutions when looking at perturbations about a uniformly moving homogeneous equilibrium state \cite{PhysRevD.31.725, PhysRevD.35.3723}. Thus, if one wishes to investigate the dynamics of a viscous relativistic fluid, it is imperative to search for a formulation where the instability and acausality issues do not arise. There have been many attempts at formulating such a theory in the past; among the most prominent examples are the M\"{u}ller-Israel-Stewart (MIS) formalism \cite{1967ZPhy198329M, ISRAEL1976310, ISRAEL1976213, ISRAEL1979341, HISCOCK1983466}, which introduces new non-hydrodynamic degrees of freedom to render the system of equations stable and causal, as well as the one by Baier et al.~\cite{Baier:2007ix} (colloquially known by the abbreviation ``BRSSS"), which introduces second order terms that (after a re-summation~\cite{Romatschke:2017ejr}) lead to a (linearly) stable and causal theory.

In recent years, there has been a renewed effort to answer the question of whether it is possible to have a self-consistent formulation of first order relativistic viscous hydrodynamics which is stable and causal, without the need to include non-hydrodynamic degrees of freedom or explicit higher derivative corrections. This has culminated in the development of the Bemfica-Disconzi-Noronha-Kovtun (BDNK) formulation of first order relativistic hydrodynamics \cite{Bemfica:2017wps, Kovtun:2019hdm, Bemfica:2019knx, Hoult:2020eho, Bemfica:2020zjp}. The novel aspect of the BDNK formulation is to take advantage of a field-redefinition freedom that exists in hydrodynamics in order to write down stable and causal hydrodynamic equations. This freedom exists because the hydrodynamic variables used in writing down the constitutive relations do not have a fundamental definition out of equilibrium.\footnote{This redefinition freedom exists even in equilibrium if the fluid is on a curved background~\cite{Kovtun:2022vas}.} Differing choices for definitions of the hydrodynamic variables are referred to in the literature as different``hydrodynamic frames."

In BDNK hydrodynamics, one works with constitutive relations written in a general hydrodynamic frame, which are less restrictive compared to formulations in the hydrodynamic frames used by Landau-Lifshitz and Eckart. In practice, this amounts to writing down every possible first-order term allowed by symmetries in the constitutive relations, which leads to the inclusion of several additional transport parameters in the equations of motion. These parameters primarily act as UV regulators for the theory, and lead to a set of stable and causal hydrodynamic equations. Of course, the number of physical transport coefficients in the theory remains unaffected even when working in a general hydrodynamic frame. Given that hydrodynamics is a classical effective field theory, the BDNK formulation exploits the inherent field redefinition freedom to its fullest to arrive at a desired stable and causal framework.\footnote{See, for instance, \cite{Bhattacharyya:2023srn, Bhattacharyya:2024tfj} for an interesting discussion of the relation between the BDNK formalism and standard Landau frame hydrodynamics. As it turns out, the first order BDNK constitutive relations when written in the Landau frame after performing a frame transformation involve all orders in derivatives. This can equivalently be expressed in an MIS-like formulation, with several new non-hydrodynamic degrees of freedom appearing that obey relaxation-type equations.} The interested reader may refer to e.g.~\cite{Bantilan:2022ech, Pandya:2022sff, Bea:2023rru, Freistuhler:2021lla} for numerical studies of relativistic viscous hydrodynamics in the BDNK formulation. 

In the present paper, our aim is to apply the BDNK procedure to first order relativistic superfluid hydrodynamics~\cite{PhysRevD.45.4536,Herzog:2011ec,Bhattacharya:2011eea,Bhattacharya:2011tra}. The system of interest carries a global $U(1)$ symmetry that gets spontaneously broken in the superfluid state. Thus, apart from the conservation of energy-momentum and the $U(1)$ charge, we also get equations governing the dynamics of the Goldstone mode associated with the spontaneous breaking of $U(1)$. It turns out that the Landau-Lifshitz-Clark-Putterman (LLCP) frame~\cite{Landau-Lifshitz, ClarkThesis, putterman1974}, often used to study relativistic viscous superfluid hydrodynamics, is also mired with problems of acausality and instability. One may then wonder whether working in a general hydrodynamic frame can cure these problems for superfluid hydrodynamics, as it does for a charged relativistic fluid with the $U(1)$ symmetry intact \cite{Hoult:2020eho, Bemfica:2020zjp, Abboud:2023hos}. Working in analytically tractable limits, our results suggest this to be the case indeed. This finding gives further credibility to the already well-established BDNK paradigm of formulating hydrodynamics in general frames, now including systems with a spontaneously broken $U(1)$ symmetry, and allows one to work with dynamical equations under which the equilibrium state is stable and perturbations propagate causally. To illustrate the procedure we work with conformal superfluids, though the same techniques will go through for a superfluid with no conformal symmetry.

This paper is organized as follows. In section \ref{sec:constitutive}, we give a detailed overview of first order relativistic superfluid hydrodynamics. We elucidate the state of thermal equilibrium, present the general first order constitutive relations when out of equilibrium, discuss frame transformations and frame invariants, and finally introduce the LLCP frame. Then, in section \ref{sec:basics}, we take a closer look at the dynamics in the LLCP frame and argue that it is unstable and acausal. Next, in section \ref{sec:new_material}, working in a general hydrodynamic frame, we enunciate the criteria transport parameters must satisfy to ensure linear stability of the equilibrium state (subsection \ref{gen_stability}), followed by laying out the criteria for linear causality of propagation for perturbations about the equilibrium state (subsection \ref{gen_causality}). We conclude with a discussion and outlook in section \ref{sec:discussion}. Appendix \ref{further_LLCP} contains additional details about the LLCP frame and the physical transport coefficients, while appendix \ref{app:Routh-Hurwitz} summarizes the mathematical criteria that need to be imposed for ensuring stability and causality of the equilibrium state. In appendix \ref{mod_Landau_frame}, we examine the ``superfluid Landau frame'' of \cite{Arean:2023nnn}, which we dub the \emph{modified Landau frame}, and argue that the dynamics in this frame too is acausal and unstable, as in the LLCP frame. 

\paragraph{Conventions:} The metric signature is mostly positive. The spacetime dimensionality is $D = d+1$. Greek indices denote time and space i.e.\ $\m, \n, \ldots = 0,1,2,\ldots d$, while Latin indices are purely spatial i.e.\ $i,j,\ldots = 1,2,\ldots d$. Boldface symbols denote spatial vectors. Symmetrization and antisymmetrization of indices is done via
\begin{equation*}
A_{(\m} B_{\n)} \equiv \half\left(A_\m B_\n + A_\n B_\m\right), \quad A_{[\m} B_{\n]} \equiv \half\left(A_\m B_\n - A_\n B_\m\right).
\end{equation*} 
A dot $``\cdot"$ between two quantities denotes contraction of spacetime indices e.g.~$a{\cdot}b \equiv a^\m b_\m = a_\m b^\m$.

\section{An overview of relativistic superfluid hydrodynamics}
\label{sec:constitutive}
We begin our discussion with an exposition of relativistic superfluid hydrodynamics as the hydrodynamic description for a Poincar\'{e} invariant system with a global $U(1)$ symmetry which undergoes spontaneous breaking. Noether's theorem for such a system implies the conservation of energy-momentum and the $U(1)$ charge density. In the hydrodynamic regime, where one observes the system on macroscopically long time and length scales, much larger than the microscopic scales associated with local equilibriation, the dynamics is governed by these conserved densities and their relaxation \cite{Landau-Lifshitz, Kovtun:2012rj}. 
However, before delving deeper into the dynamics of superfluids in a general out of equilibrium state, we first take a careful look at the state of thermal equilibrium - see section \ref{super_equilibrium}. Since a generic out of equilibrium state relaxes to equilibrium when the fluctuations vanish, the equilibrium state can be used to find important constraints on the constitutive relations for the out of equilibrium state, as we elucidate in section \ref{constitutive_rels}. Section \ref{frames} then discusses hydrodynamic frame transformations for superfluids, while section \ref{LLCPframe} specifically focuses on the frequently used Landau-Lifshitz-Clark-Putterman (LLCP) frame.  

\subsection{Thermal equilibrium}
\label{super_equilibrium}
In thermal equilibrium, a fluid endowed with a global $U(1)$ symmetry, often referred to as a ``charged'' fluid, is well described by the grand canonical ensemble characterized by the temperature $T$, chemical potential $\mu$, and the fluid velocity $u^\m$, normalized such that $u^\m u_\m = -1$. The physical properties of the fluid in equilibrium, such as the conserved currents associated with various symmetries, as well as their correlation functions, can be obtained from an equilibrium generating functional \cite{Banerjee:2012iz, Jensen:2012jh}, which is a functional of the background sources. To be precise, the background metric $g_{\m\n}$ and the background gauge field $A_\m$ source the fluid energy-momentum tensor $T^{\m\n}$ and the $U(1)$ current $J^\m$, respectively. The equilibrium state is further characterized by the presence of a timelike Killing vector $K^\m$, which specifies the direction of time on the given background, and a gauge function $\L_K$. The sources in equilibrium satisfy the conditions
\begin{subequations}
\label{Lie_conds}
\begin{align}
&\pounds_K g_{\m\n} = 0, \label{Lie_met}\\ 
&\pounds_K A_\m + \del_\m \L_K = 0, \label{Lie_gauge}
\end{align}
\end{subequations}
where $\pounds_K$ denotes the Lie derivative with respect to $K^\m$. Under a gauge transformation $A_\m \rightarrow A_\m + \del_\m \l$, the Killing vector $K^\m$ is invariant, whereas the gauge function transforms as $\L_K \rightarrow \L_K + \d_\l \L_K$, with $\d_\l \L_K = - \pounds_K \l$.\footnote{This leads to the equilibrium conditions in eq.~\eqref{Lie_conds} being gauge invariant.} 
By virtue of this, the Killing vector and the gauge function furnish a gauge invariant definition for the temperature, chemical potential and fluid velocity in equilibrium, given by
\be
\label{thermo_defs}
T \equiv \frac{T_0}{\sqrt{-K^2}}\, , \quad \m \equiv \frac{K^\m A_\m+\L_K}{\sqrt{-K^2}}\, , \quad u^\m \equiv \frac{K^\m}{\sqrt{-K^2}}\, ,
\ee
where $T_0$ is a normalization constant setting the equilibrium temperature scale. 

When the global $U(1)$ symmetry gets spontaneously broken, the system enters a superfluid phase. Microscopically, the spontaneous breakdown happens because an operator charged under the symmetry condenses and acquires a non-vanishing expectation value in the equilibrium state. The phase of the condensed operator, denoted $\varphi$, emerges as an additional degree of freedom for the system - it is the Goldstone mode associated with the spontaneous breaking of the global $U(1)$ symmetry.\footnote{Throughout the paper, we work in the regime far removed from the superfluid critical point, decoupling the gapped ``amplitude mode'' \cite{Pekker-Varma, Donos:2022xfd} from the rest of the dynamics.} 
Under the gauge transformation $A_\m \rightarrow A_\m + \del_\m \l$, it undergoes a shift via $\vp \rightarrow \vp + \l$. In thermal equilibrium, we demand that the Goldstone $\vp$ is invariant under a combination of diffeomorphisms generated by $K^\m$ and gauge transformations generated by $\L_K$, via
\be
\pounds_K \vp + \L_K = 0.
\label{Lie_Gold}
\ee
The condition \eq{Lie_Gold} supplements the previously stated equilibrium conditions \eq{Lie_conds}. 

We now introduce the gauge invariant combination $\xi_\m \equiv - \del_\m \vp + A_\m$, which by virtue of its definition satisfies
\be
\label{xi_Eq}
\N_\m \xi_\n - \N_\n \xi_\m = F_{\mu\nu}\, ,
\ee
with $F_{\mu\nu} \equiv \del_\m A_\n - \del_\n A_\m$ being the field strength tensor for the gauge field $A_\m$. 
It is important to mention here that with regards to derivative counting, one takes the Goldstone mode $\vp$ to be $\mc{O}(\del^{-1})$, implying that its derivative $\xi^\m$ is an $\mc{O}(1)$ quantity. Notice that $\xi^\m$ has a well-defined microscopic definition, just as the energy-momentum tensor and the $U(1)$ current. In what follows, we will treat $\xi^\m$ as an independent (non-conserved) current,\footnote{This stands in contrast with an alternate treatment of $\xi^\m$ in parts of the literature e.g.~\cite{Herzog:2011ec,Bhattacharya:2011eea,Bhattacharya:2011tra,Jensen:2012jh,Bhattacharyya:2012xi}, where it is treated as a hydrodynamical variable characterizing the equilibrium state at par with $T, \m$ etc.} at par with $T^{\m\n}$ and $J^\m$, whose evolution is constrained by eq.\ \eqref{xi_Eq}.

Utilizing the definitions in \eq{thermo_defs}, the equilibrium condition \eq{Lie_Gold} implies that
\be
u^\m \xi_\m = \m.
\label{Josephson}
\ee
In other words, the chemical potential is the time derivative of the Goldstone mode. This relation is known as the \emph{Josephson equation}. In our construction, it naturally arises for the equilibrium state without making any additional assumptions.
It is now convenient to introduce the decomposition $\xi^\m = - \m u^\m + \z^\m$, with $\z^\m$ being the spacelike projection of $\xi^\m$ orthogonal to the fluid velocity i.e.\ $\z^\m \equiv \D^{\m\n} \xi_\n$, with $\D_{\m\n} \equiv g_{\m\n} + u_\m u_\n$ being the projector orthogonal to $u^\m$. This decomposition has the virtue of automatically satisfying the Josephson condition eq.~\eqref{Josephson}. The equilibrium state in the superfluid phase is thus characterized in terms of $T, \m, u^\m$ and $\z^\m$.

The equilibrium generating functional has the form \cite{Banerjee:2012iz, Jensen:2012jh, Bhattacharyya:2012xi}
\begin{equation}
W[g,A,\vp] = - \int d^{d+1}x \sqrt{-g}\, {\cal F}(g, A, \vp)\, , \label{gen_func}
\end{equation}
where the free energy density $\mc{F}$ is a local function of the sources and the Goldstone mode. As mentioned earlier, the metric acts as the source for the energy-momentum tensor $T^{\m\n}$, whereas the gauge field sources the $U(1)$ current $J^\m$. Varying the generating functional therefore gives
\be
\label{def_var_W}
\d W = \int d^{d+1}x \sqrt{-g} \left( \half \, T^{\m\n} \d g_{\m\n} + J^\m \d A_\m + E_\vp \d\vp \right), 
\ee
with $E_\vp = 0$ being the equation of motion for $\vp$. The generating functional thus acts as an effective action for the Goldstone mode, in the presence of a background metric and gauge field. Under a diffeomorphism generated by a vector field $\chi^\m$, the variation of the generating functional is given by
\be 
\d W = \int d^{d+1}x \sqrt{-g} \left( \half \, T^{\m\n} (\pounds_\chi g)_{\m\n} +  J^\m (\pounds_\chi A)_\m + E_\vp \pounds_\chi \vp\right).
\ee
Demanding invariance of the generating functional under diffeomorphisms leads to the condition
\be
\label{div_T_E}
\N_\m T^{\m\n} = F^{\n\l} J_\l - (\N_\m J^\m) A^\n + E_\vp \del^\n \vp.
\ee
Similarly, under the gauge transformation $A_\m \rightarrow A_\m + \del_\m \l$, the change in the generating functional is
\be
\d W = \int d^{d+1}x \sqrt{-g} \left( J^\m \del_\m \l + \l E_\vp\right). 
\ee
Demanding the gauge invariance of the generating functional yields
\be
\label{div_J_E}
\N_\m J^\m = E_\vp.
\ee
Thus, it is only when the Goldstone mode is onshell, $E_\vp = 0$, that \eq{div_T_E} and \eq{div_J_E} give rise to the standard energy-momentum and current conservation equations
\begin{subequations}
\label{hydro_eqs}
\begin{align}
\N_\m T^{\m\n} &= F^{\n\l} J_{\l}\, ,\label{hydro_eq_1}\\
\N_\m J^\m &= 0. \label{hydro_eq_2}
\end{align}
\end{subequations}
Put differently, gauge invariance implies that the equation of motion of the Goldstone mode can be written as \eq{hydro_eq_2}. A given equilibrium state thus corresponds to given background sources $g_{\m\n}, A_\m$ satisfying \eq{Lie_conds}, along with a configuration of $\vp$ which satisfies its equation of motion and the condition \eq{Lie_Gold}.

Now, as the free energy density satisfies locality, it admits a derivative expansion in terms of the hydrodynamic variables. At the leading order i.e.~with zero derivatives, it simply equals minus the pressure $P$ as a function of the diffeomorphism and gauge invariant scalars $T, \m, \z$ which characterize the equilibrium state, with $\z \equiv \sqrt{\z^\m \z_\m}$ being the norm of $\z^\m$. Thus we have
\be
\label{gen_func_1}
W[g, A, \vp] = \int d^{d+1}x \sqrt{-g} \, P(T, \m, \z) + \mc{O}(\del).
\ee
The pressure $P$ satisfies the Gibbs-Duhem relation
\be
\label{Gibbs-Duhem}
dP = s \, dT + \rho \,d\m + \tilde{\rho} \, d\z\, ,
\ee
with $s$ being the entropy density. Using eq.~\eqref{def_var_W}, the energy-momentum tensor and current that follow from the variation of the generating functional \eq{gen_func_1} are given by
\begin{subequations}\label{ideal_currs}
\begin{align}
T^{\m\n} &= \e u^\m u^\n + \tilde{\e} z^\m z^\n + P \tilde{\D}^{\m\n} + 2 \m \tilde{\rho} u^{(\m} z^{\n)}\, ,\label{ideal_stress}\\
J^\m &= \rho u^\m + \tilde{\rho} z^\m \, , \label{ideal_curr}
\end{align}
\end{subequations}
where $z^\m \equiv \z^\m/\z$ is the unit spacelike vector along $\z^\m$, and $\tilde{\D}_{\m\n} \equiv g_{\m\n} + u_\m u_\n - z_\m z_\n$ is the projector orthogonal to both $u^\m$ and $z^\m$. Also, we have made use of the definitions
\be
\label{etedefs}
\begin{split}
\e &\equiv - P + T s + \m \rho\, ,\\
\tilde{\e} &\equiv P - \z \tilde{\rho}\, .
\end{split}
\ee
Due to the diffeomorphism and gauge invariance of \eq{gen_func_1}, the energy-momentum tensor and current in \eq{ideal_currs} automatically satisfy the conservation equations \eqref{hydro_eqs}.\footnote{Instead of \eq{gen_func_1}, one can equivalently work with the generating functional
$$W[g, A, \vp] = \int d^{d+1}x \sqrt{-g} \, P(T,\m,\xi) + \mc{O}(\del)\, ,$$
where $\xi \equiv \sqrt{-\xi^\m \xi_\m}$, with the Gibbs-Duhem relation given by $dP = s\, dT + n \, d\m + \tilde{n} \, d\xi$,  and $n, \f{\m}{\xi} \tilde{n}$ respectively identified as the charge densities associated with the normal and superfluid components \cite{Herzog:2008he}. The equilibrium energy-momentum tensor and current now have the form
\begin{align*}
T^{\m\n} &= \e_n u^\m u^\n + P \D^{\m\n} + \e_s \f{\xi^\m \xi^\n}{\xi^2}\, ,\\
J^\m &= n u^\m - \tilde{n} \f{\xi^\m}{\xi}\, ,
\end{align*}
with $\e_n \equiv - P + T s + \m n$ being the energy density associated with the normal component, and $\e_s \equiv \xi \tilde{n}$ proportional to the energy density of the superfluid component, $\f{\m^2}{\xi^2} \e_s$. The various quantities in the two equivalent descriptions are related via
$$\e = \e_n + \f{\m^2}{\xi^2} \e_s\, , \quad \tilde{\e} = P + \f{\z^2}{\xi^2} \e_s\, , \quad \rho = n + \f{\m}{\xi} \tilde{n}\, , \quad \tilde{\rho} = - \f{\z}{\xi} \tilde{n}.$$ 
Thus the quantities $\e, \rho$ in \eq{Gibbs-Duhem} \& \eq{ideal_currs} depict the total energy and charge densities of the state.
} The expressions in eq.~\eqref{ideal_currs} that arise at the zeroth order in derivatives are also referred to as the ideal/perfect superfluid constitutive relations. \\

\noindent \textit{\underline{Equilibrium relations}:} By virtue of the equilibrium conditions \eq{Lie_conds} and \eq{Lie_Gold}, we obtain several relations between the thermodynamic quantities and their derivatives. For instance, the conditions \eq{Lie_conds} along with the definitions \eq{thermo_defs} imply that in thermal equilibrium one has
\begin{subequations}
\label{eqconds}
\begin{align}
\N_\l T &= -T a_\l\, ,\label{eqcond1}\\
\N_\l \m &= - \m a_\l + E_\l\, ,\label{eqcond2}\\
\N_\m u_\n &= - u_\m a_\n - \half \e_{\m\n\rho\s} u^\rho \O^\s\, , \label{eqcond3}
\end{align}
\end{subequations}
where $a^\m \equiv u^\n \N_\n u^\m$ is the acceleration four-vector, $E^\m \equiv F^{\m\n} u_\n$ is the (background) electric field, and $\O^\m \equiv \e^{\m\n\rho\s} u_\n \N_\rho u_\s$ is the vorticity four-vector -- all of which are orthogonal to $u^\m$. Note that in our derivative counting scheme $A_\m \sim \mc{O}(1)$, and hence $E^\m \sim \mc{O}(\del)$.

Additionally, \eq{Lie_Gold} along with \eq{Lie_conds} implies that $\pounds_K \xi_\m = 0$, which can equivalently be expressed as $\pounds_K \z_\m = 0$. This leads to
\be
u^\l \N_\l z_\m  = u_\m \,z\cdot a  - \f{1}{\z}\, z_\m u^\l \N_\l \z + \half \,\e_{\m\rho\a\b} z^\rho u^\a \O^\b.
\label{eqcond4}
\ee
Another set of equilibrium conditions is $\pounds_K \xi^2 = \pounds_K \z^2 = 0$. It is important to realize that the relations \eq{eqconds} and \eq{eqcond4} follow purely from equilibrium conditions and the definitions for the hydrodynamic variables, and do not depend upon the form of the generating functional. They are true to any order in the derivative expansion. \\

\noindent \textit{\underline{Generating functional at $\mc{O}(\del)$}:} 
\Eq{gen_func_1} gives the form of the equilibrium generating functional at the leading order in the derivative expansion i.e.\ without any derivatives. The only term that appears is the pressure $P(T,\m,\z)$. At the first order in derivatives for a parity-preserving system, there are eleven diffeomorphism and gauge invariant scalars constructed using the hydrodynamic variables that can generically contribute to the generating functional, provided time reversal invariance is not a symmetry. If it is, as will be the case for us, then one does not get any first order contributions, as further discussed below. For the sake of completeness, we state the eleven scalars, which are\footnote{For parity violating systems, one can have two additional scalars at $\mc{O}(\del)$ in the generating functional: $B\cdot z$ and $\O\cdot z$, where $B^\m \equiv \half \e^{\m\n\rho\s} u_\n F_{\rho\s}$ is the magnetic field. However, in the present work, our focus is on parity preserving systems, thereby disallowing such terms.}
\be
\begin{split}
u^\a \N_\a T\, ,\quad &u^\a \N_\a \m\, ,\quad u^\a \N_\a \z\, ,\quad z^\a \N_\a T\, ,\quad z^\a \N_\a \m\, ,\quad z^\a \N_\a \z\, , \\
&\N\cdot u\, ,\quad \N\cdot z\, ,\quad z\cdot a\, ,\quad z^\a z^\b \N_\a u_\b\, ,\quad E\cdot z\,.
\end{split}
\ee
However, the equilibrium conditions \eq{eqcond1} - \eq{eqcond3} along with $\pounds_K \z^2 = 0$ imply that $u^\a \N_\a T\, ,u^\a \N_\a \m\, , u^\a \N_\a \z\,, z^\a z^\b \N_\a u_\b$ and $\N\cdot u$ vanish identically. Further, contracting \eq{eqcond1} and \eq{eqcond2} with $z^\a$ provides two more relations between the remaining six non-vanishing $\mc{O}(\del)$ scalars, given by
\be
\label{eqcond5}
\begin{split}
&z^\a \N_\a T + T \,z\cdot a = 0,\\
&z^\a \N_\a \m + \m\, z\cdot a - E\cdot z = 0.
\end{split}
\ee
An additional constraint is provided by the equation of motion for the dynamical Goldstone mode $\vp$, which is identical to \eq{hydro_eq_2}, with the constitutive relation for $J^\m$ given at the zeroth order in derivatives by \eq{ideal_curr}. This leads to
\be
\label{eqcond6}
\tilde{\rho} \, \N\cdot z + \f{\del\tilde{\rho}}{\del T} \, z^\a \N_\a T + \f{\del\tilde{\rho}}{\del \m} \, z^\a \N_\a \m + \f{\del\tilde{\rho}}{\del \z} \, z^\a \N_\a \z = 0. 
\ee
Therefore, at $\mc{O}(\del)$, there are only three independent non-vanishing diffeomorphism and gauge invariant scalars that can appear in the generating functional.\footnote{The remaining equations in thermal equilibrium, eqs.~\eqref{xi_Eq}, \eqref{hydro_eq_1}, and \eqref{eqcond4} do not provide any additional constraints on the $\mc{O}(\del)$ scalars beyond what have already been obtained above.} These can be chosen to be $z^\a \N_\a T, z^\a \N_\a \m$ and $z^\a \N_\a \z$. Varying the generating functional with respect to the sources will now give rise to equilibrium constitutive relations for $T^{\m\n}$ and $J^\m$ with $\mc{O}(\del)$ hydrostatic correction terms beyond the perfect superfluid constitutive relations of \eqref{ideal_currs}. 

An important point to note is that all six of the non-vanishing $\mc{O}(\del)$ scalars in thermal equilibrium are odd under time reversal, since $z^\m$ is odd under time reversal. Thus, if the underlying microscopic theory respects time reversal invariance, then none of these $\mc{O}(\del)$ scalars can contribute to the generating functional. In other words, for a time reversal invariant system, the perfect superfluid constitutive relations \eq{ideal_currs} do not receive any $\mc{O}(\del)$ correction terms from the equilibrium generating functional. As mentioned above, in the present work, we indeed consider the underlying microscopic theory to respect time reversal invariance, and therefore any $\mc{O}(\del)$ corrections to the constitutive relations \eq{ideal_currs} should be out-of-equilibrium contributions, as we now discuss next in section \ref{constitutive_rels}.\footnote{See ref.\ \cite{Bhattacharyya:2012xi} for a discussion of relativistic superfluid hydrodynamics when time reversal invariance is allowed to be broken.}

\subsection{Out of equilibrium constitutive relations}
\label{constitutive_rels}
Out of equilibrium, in the hydrodynamic regime, the variables $T, \m, u^\m$ and $\z^\m$ become slowly varying functions of space and time. Beyond the equilibrium constitutive relations, $T^{\m\n}, J^\m$ and $\xi^\m$ now acquire derivative corrections which can be dissipative. One can organize the derivative corrections in a systematic manner using the symmetries of the system. 
Given the unit timelike vector $u^\m$ and the unit spacelike vector $z^\m$, we can decompose $T^{\m\n}, J^\m$ and $\xi^\m$ in terms of objects that transform like scalars, vectors and tensors under an $SO(2)$ transformation. This decomposition takes the form
\begin{subequations}
\label{eq:constit_decomp}
\begin{align}
T^{\mu\nu} &= \ET u^\mu u^\nu + \VT z^\mu z^\nu + \PT \tD^{\mu\nu} + 2 \, \UT u^{(\mu} z^{\nu)} + 2\QT^{(\mu} u^{\nu)} + 2 \RT^{(\mu} z^{\nu)} + \TT^{\mu\nu}\, ,\label{stress_gen}\\
J^\mu &= \NJ u^\mu + \SJ z^\mu + \JJ^\mu\, ,\label{curr_gen}\\
\xi^\mu &= -\MX u^\mu + \ZX z^\mu + \XX^\mu\, ,\label{xi_gen}
\end{align}
\end{subequations}
where the projector $\tilde{\D}_{\m\n} \equiv g_{\m\n} + u_\m u_\n - z_\m z_\n$ projects orthogonal to the plane containing $u^\m$ and $z^\m$.
The various quantities appearing in the decomposition above are defined via
\be
\label{inverse_decomp}
\begin{split}
&\ET \equiv u_\mu u_\nu T^{\mu\nu}, \quad \VT \equiv z_\mu z_\nu T^{\mu\nu}, \quad \PT \equiv \frac{1}{\lr{d-1}} \tD^{\mu\nu} T_{\mu\nu}, \quad \UT \equiv - u_\mu z_\nu T^{\mu\nu},\\
&\NJ \equiv - u^\mu J_\mu,\quad\,\,\, \SJ \equiv z^\mu J_\mu, \qquad\,\, {\cal M} \equiv u^\mu \xi_\mu, \qquad\qquad\quad {\cal Z} \equiv z^\mu \xi_\mu,\\
&\QT^\alpha \equiv -\tD^{\alpha\mu} u^\nu T_{\mu\nu}, \qquad \RT^{\alpha} \equiv \tD^{\alpha\mu} z^\nu T_{\mu\nu}, \qquad \JJ^\mu \equiv \tD^{\mu\nu} J_{\nu}, \qquad \XX^\mu \equiv \tD^{\mu\nu}\xi_\nu,\\
&\TT^{\mu\nu} \equiv \f{1}{2} \lr{\tD^{\mu\alpha} \tD^{\nu\beta} + \tD^{\mu\beta} \tD^{\nu\alpha} - \frac{2}{\lr{d-1}} \tD^{\mu\nu} \tD^{\alpha\beta}}T_{\a\b}.
\end{split}
\ee
This ensures that the decomposition in eq.\ \eqref{eq:constit_decomp} is invertible. Note that $\QT^\m, \RT^\m, \JJ^\m, \XX^\m$ and $\TT^{\m\n}$ are all orthogonal to both $u^\m$ and $z^\m$, with $\TT^{\m\n}$ further being traceless.

For the superfluid state in thermal equilibrium, the constitutive relations eq.\ \eqref{eq:constit_decomp} take a particularly simple form. Comparing with eq.~\eqref{ideal_currs}, one has
\be
\mc{E}_{\rm eq} = \epsilon, \,\, \mc{V}_{\rm eq} = \tilde{\epsilon}, \,\, \mc{P}_{\rm eq} = P, \,\, \mc{U}_{\rm eq} = \m \tilde{\rho}, \,\, \mc{N}_{\rm eq} = \rho, \,\, \mc{S}_{\rm eq} = \tilde{\rho}, \,\, \mc{M}_{\rm eq} = \m, \,\, \mc{Z}_{\rm eq} = \z,
\label{ideal_currs_bulk}
\ee 
while the rest vanish, where the subscript ``eq" denotes the equilibrium state.

For the out of equilibrium state, at first order in derivatives, the constitutive relations read
\begin{subequations}
\label{eq:most_gen_decomp}
\begin{align}
T^{\m\n} &= \lr{\e + \ET_{1}} u^\m u^\n + \lr{\te + \VT_1}z^\m z^\n + \lr{P + \PT_1} \tD^{\m\n} + 2 \lr{\mu \tilde{\rho} + \UT_1}u^{(\m} z^{\n)} \nonumber \\
&\quad + 2 \QT_{1}^{(\m} u^{\n)} + 2 \RT_{1}^{(\m} z^{\n)} + \TT_{1}^{\m\n}\, ,\label{g_decomp_1}\\
J^\m &= \lr{\rho + \NJ_{1}} u^\m + \lr{\tilde{\rho}+ \SJ_1} z^\m + \JJ_{1}^\m \, ,\label{g_decomp_2}\\
\xi^\m &= - \lr{\m + \MX_1} u^\m + \lr{\zeta + \ZX_1}z^\mu + \XX_1^\mu\, . \label{g_decomp_3}
\end{align}
\end{subequations}
In the decomposition above, the subscript $``1"$ signifies a one derivative correction term arising from the slowly varying hydrodynamic variables $T, \m, u^\m$ and $\z^\m$ in the out of equilibrium state. In principle, two types of contributions are possible: those that vanish in the limit of thermal equilibrium, and those that do not. To maintain consistency with thermal equilibrium, the contributions that do not vanish in the equilibrium limit should arise from the equilibrium generating functional. Given that we are interested in systems which at the microscopic level preserve time-reversal invariance, there are no first order corrections in the constitutive relations eq.\ \eqref{eq:most_gen_decomp} that arise from an equilibrium generating functional and hence survive in the equilibrium limit, as was discussed in more detail in section \ref{super_equilibrium}. Thus, all one derivative terms in eq.\ \eqref{eq:most_gen_decomp} consist of scalars, vectors and tensors that vanish when thermal equilibrium is restored, which we enumerate below. In the subsequent analysis, we will turn off the external gauge field i.e.~set $A_\m = 0$.

\noindent$\bullet$ \textit{\underline{Scalar data}:} Using the results of section \ref{super_equilibrium}, one can construct eight independent one derivative scalars that vanish in thermal equilibrium: the five scalars that identically vanished, as well as the linear combinations in eqs.~\eqref{eqcond5},~\eqref{eqcond6}. With slight rewriting, they take the form
\be
\begin{split}
&\f{1}{T} \, u^\a \N_\a T\, ,\quad u^\a \N_\a \left(\f{\m}{T}\right)\, ,\quad u^\a \N_\a \left(\f{\z}{T}\right)\, ,\quad z^\a z^\b \N_\a u_\b\, ,\quad \N\cdot u\, ,\\
&\f{1}{T} \, z^\a \N_\a T + z\cdot a \, ,\quad
z^\a \N_\a \left(\f{\m}{T}\right)\, ,\quad \f{1}{T^{d-1}} \, \D^{\a\b} \N_\a \left(\f{\tilde{\rho} \, z_\b}{T}\right).
\end{split}
\label{scalar_data}
\ee
The above eight scalars are therefore allowed to appear as out of equilibrium derivative corrections at first order in the constitutive relations eq.~\eqref{eq:most_gen_decomp}. 

\noindent$\bullet$ \textit{\underline{Vector data}:} We are now interested in enumerating one derivative vectors that can contribute to the out of equilibrium constitutive relations. These vectors must be transverse to the plane containing  $u^\m$ and $z^\m$, and must vanish in thermal equilibrium. To begin with, there are eight transverse vectors with one derivative, given by
\be
\label{org_v_data}
\begin{split}
&\tilde{\D}^{\m\n} \N_\n T\, ,\quad \tilde{\D}^{\m\n} \N_\n \m\, ,\quad \tilde{\D}^{\m\n} \N_\n \z\, ,\quad \tilde{\D}^{\m\n} a_\n\, ,\\
&\tilde{\D}^{\m\n} u^\rho\N_\rho z_\n\, ,\quad \tilde{\D}^{\m\n} z^\rho \N_\rho z_\n\, ,\quad \tilde{\D}^{\m\n} z^\rho \N_\rho u_\n\, ,\quad \tilde{\D}^{\m\n} z^\rho \N_\n u_\rho\, .
\end{split}
\ee
The equilibrium conditions \eq{eqconds} and \eq{eqcond4} lead to the vanishing of four combinations made up of the eight vectors above. These combinations are
\be
\label{v_van_1}
\begin{split}
&\tilde{\D}^{\m\n}\left(\N_\n T + T a_\n\right),\quad \tilde{\D}^{\m\n}\left(\N_\n \m + \m a_\n\right), \\
&\tilde{\D}^{\m\n} \left(z^\rho \N_\rho u_\n - u^\rho \N_\rho z_\n\right),\quad \tilde{\D}^{\m\n}\left(z^\rho \N_\n u_\rho + u^\rho \N_\rho z_\n\right).
\end{split}
\ee
Additionally, the transverse vector projection that arises from the condition \eq{xi_Eq} i.e.\ $\tilde{\D}^{\rho\n} \xi^\m \left(\N_\m \xi_\n - \N_\n \xi_\m\right) = 0$ leads to the vanishing of the combination
\be
\label{v_van_2}
\tilde{\D}^{\m\n} \left(z^\rho \N_\rho z_\n - 2\m z^\rho \N_\rho u_\n - \z \N_\n \z\right)
\ee
in thermal equilibrium. The transverse projection of the stress tensor conservation \eq{hydro_eq_1} does not lead to the vanishing of any additional vector in equilibrium beyond the ones mentioned above. Eqs.~\eqref{v_van_1} and \eqref{v_van_2} thus provide the exhaustive list of one derivative transverse vectors that can appear in the out of equilibrium constitutive relations.

\noindent$\bullet$ \textit{\underline{Tensor data}:} Finally, let us look at the independent symmetric transverse traceless tensors at the first order in derivatives that can contribute to the constitutive relations. \textit{A priori}, there are two possibilities, denoted by
\begin{subequations}
\label{v_tens}
\begin{align}
\s^{\m\n}_u &\equiv \left(\tilde{\D}^{\m\a} \tilde{\D}^{\n\b} + \tilde{\D}^{\m\b} \tilde{\D}^{\n\a} - \f{2}{d-1}\,\tilde{\D}^{\m\n} \tilde{\D}^{\a\b}\right) \N_\a u_\b\, ,\label{tens_1}\\
\s^{\m\n}_z &\equiv \left(\tilde{\D}^{\m\a} \tilde{\D}^{\n\b} + \tilde{\D}^{\m\b} \tilde{\D}^{\n\a} - \f{2}{d-1}\,\tilde{\D}^{\m\n} \tilde{\D}^{\a\b}\right) \N_\a z_\b\, .\label{tens_2}
\end{align}
\end{subequations}
Out of these, by virtue of the equilibrium condition \eq{eqconds}, $\s^{\m\n}_u$ vanishes identically in thermal equilibrium, and is therefore the only allowed one derivative tensor to appear in the out of equilibrium constitutive relations eq.~\eqref{eq:most_gen_decomp}.

To summarize, there are eights scalars, eq.\ \eqref{scalar_data}, five transverse vectors, eqs.\ \eqref{v_van_1} and \eqref{v_van_2},  and one symmetric, transverse, traceless tensor, eq.\ \eqref{tens_1}, that comprise the independent one derivative terms that vanish in equilibrium and are hence allowed to appear as derivative corrections in the out of equilibrium constitutive relations eq.\ \eqref{eq:most_gen_decomp}.

It is important to note that the hydrodynamic equations satisfied by the ideal constitutive relations of eq.~\eqref{ideal_currs} provide additional ``onshell" constraints on one derivative scalar and vector quantities, thereby leading to a reduction in the number of possible such terms compared to the above offshell counting. Performing this onshell reduction is the usual approach taken in hydrodynamic theories; see for instance \cite{Kovtun:2012rj}. However, to understand the causality and stability properties of the hydrodynamic equations, it is imperative to allow for all possible offshell terms to appear, without imposing any onshell relations amongst them. Such relations and the ensuing reduction in first order terms fundamentally changes the mathematical nature of the hydrodynamic equations as a set of coupled partial differential equations governing the time evolution of the out of equilibrium state, and hence affect the causality and stability aspects. We will thus not make use of any onshell constraints on the allowed one derivative terms, and will therefore be working with the most general set of hydrodynamic equations.\\

\noindent \textbf{\underline{The conformal limit}:}\\
As mentioned in the introduction (section \ref{sec:Intro}), our focus in the present work will be on conformal superfluids. The requirement of conformal invariance imposes further restrictions on the allowed structures in the constitutive relations. For one, conformal symmetry requires the invariance of the equilibrium generating functional \eq{gen_func} under a Weyl rescaling of the metric $g_{\m\n} \rightarrow e^{-2\psi(x)} g_{\m\n}$. This implies that the energy-momentum tensor and the $U(1)$ current, which are defined in terms of the variation of the generating functional \eq{def_var_W}, must have the Weyl transformation property
\be
T^{\m\n} \rightarrow e^{(d+3)\psi} \, T^{\m\n}\, , \quad J^\m \rightarrow e^{(d+1)\psi} J^\m\, .
\ee
A quantity that transforms via $O \rightarrow e^{\boldsymbol{w} \psi} O$ under the Weyl rescaling $g_{\m\n} \rightarrow e^{-2\psi} g_{\m\n}$ is said to have the Weyl weight $\boldsymbol{w}$. Thus, the Weyl weights of the energy-momentum tensor and the $U(1)$ current are $d+3$ and $d+1$ respectively. 

For the constitutive relations, the fact that the energy-momentum tensor and the $U(1)$ current have well-defined Weyl weights requires that all the allowed structures that appear in them should themselves transform homogeneously under a Weyl rescaling of the metric, and thus carry well-defined Weyl weights. For the hydrodynamic variables \eq{thermo_defs} and the thermodynamic quantities that appear in the perfect superfluid constitutive relations \eq{ideal_currs} the Weyl weights are summarized in the table below.
\begin{center}
\begin{tabular}{ |c|c|c|c|c|c|c|c|c|c|c|c|c|c|c|c| } 
 \hline
 Quantity & $T$ & $\m$ & $u^\m$ & $u_\m$ & $\xi^\m$ & $\xi_\m$ & $z^\m$ & $z_\m$ & $\z$ & $P$ & $\e$ & $\tilde{\e}$ & $s$ & $\rho$ & $\tilde{\rho}$ \\ 
 \hline
 $\boldsymbol{w}$ & 1 & 1 & 1 & $-1$ & 2 & 0 & 1 & $-1$ & 1 & $d+1$ & $d+1$ & $d+1$ & $d$ & $d$ & $d$ \\ 
 \hline
\end{tabular}
\end{center}

For the out-of-equilibrium constitutive relations, the requirement of transforming homogeneously under a Weyl rescaling imposes additional constraints on the otherwise allowed scalar, vector and tensor structures. The eight independent scalars that can appear in the constitutive relations at first order in derivatives were presented in \eq{scalar_data}. It turns out that further demanding homogeneity under a Weyl rescaling only allows for seven of them to appear in the out-of-equilibrium constitutive relations. These are
\be
\begin{split}
&\mbb{s}_1 \equiv \f{1}{T}\, u^\a \N_\a T + \f{1}{d}\, \N\cdot u\, ,\quad \mbb{s}_2 \equiv u^\a \N_\a\left(\f{\m}{T}\right),\qquad\quad\,\, \mbb{s}_3 \equiv u^\a \N_\a \left(\f{\z}{T}\right) , \\
&\mbb{s}_4 \equiv z^\a \N_\a\left(\f{\m}{T}\right), \qquad\quad\,\,\, \mbb{s}_5 \equiv \f{1}{T}\, z^\a \N_\a T + z\cdot a\, ,\qquad\,\, \mbb{s}_6 \equiv z^\a z^\b \N_\a u_\b - \f{1}{d}\, \N\cdot u\, ,\\
&\mbb{s}_7 \equiv \f{1}{T^{d-1}}\, \D^{\a\b} \N_\a\left(\f{\tilde{\rho} \, z_\b}{T}\right). 
\end{split}
\label{scalar_data_conf}
\ee
All of the seven one derivative scalars above have the Weyl weight $\boldsymbol{w} = 1$.

For one derivative transverse vectors, \eq{v_van_1} and \eq{v_van_2}, we choose the following combinations which transform homogeneously under a Weyl rescaling of the metric:
\be
\begin{split}
&\mbs{v}_1^\m \equiv \tilde{\D}^{\m\n} \left(\f{1}{T} \N_\n T + a_\n\right),\qquad \mbs{v}_2^\m \equiv \tilde{\D}^{\m\n} \N_\n\left(\f{\m}{T}\right), \\
&\mbs{v}_3^\m \equiv 2 \tilde{\D}^{\m\n} z^\l \N_{(\l} u_{\n)}\, ,\qquad\qquad\,\, \mbs{v}_4^\m \equiv \tilde{\D}^{\m\n} \left(u^\l \N_\l z_\n + z^\l \N_\n u_\l\right),\\
&\mbs{v}_5^\m \equiv \tilde{\D}^{\m\n}\left[\N_\n\left(\f{\z}{T}\right) +\f{\m}{T} \left(u^\l \N_\l z_\n + z^\l \N_\l u_\n\right) + \f{\z}{T} \left(\f{1}{T}\, \N_\n T - z^\l \N_\l z_\n\right)\right].
\end{split}
\label{vector_data_conf}
\ee
All of the transverse vectors above have the Weyl weight $\boldsymbol{w} = 2$. 

Finally, the only dissipative symmetric transverse traceless tensor at first order in derivatives, $\s^{\m\n}_{u}$, continues to be allowed in the conformal limit since it transforms homogeneously under a Weyl rescaling of the metric, with the Weyl weight $\boldsymbol{w} = 3$. 

Apart from Weyl covariance, conformal invariance requires the energy-momentum tensor eq.\ \eqref{stress_gen} to be traceless, which implies $\mc{E} - \mc{V} = (d-1) \mc{P}$. In particular, in thermal equilibrium eq.~\eqref{ideal_stress}, one has $\e - \tilde{\e} = (d-1) P$.

Given the first order data eqs.\ \eqref{scalar_data_conf}, \eqref{vector_data_conf} and \eqref{tens_1}, we can now write down the most general expressions for the various one derivative quantities that appear in the out of equilibrium constitutive relations eq.\ \eqref{eq:most_gen_decomp}. We have
\begin{align}
\ET_{1} &= \sum_{n=1}^7 \ve_n \mbb{s}_n, \quad\, 
\VT_1= \sum_{n=1}^7 \lr{\ve_n - \lr{d-1} \pi_n} \mbb{s}_n, \quad 
\PT_1 = \sum_{n=1}^7 \pi_n \mbb{s}_n, \quad \UT_1 = \sum_{n=1}^7 \varphi_n \mbb{s}_n,\nonumber \\
\NJ_1 &= \sum_{n=1}^7 \nu_n \mbb{s}_n, \quad \,
\SJ_1 = \sum_{n=1}^7 \lambda_n \mbb{s}_n, \quad\,\,
{\cal M}_1 = \sum_{n=1}^7 \alpha_n \mbb{s}_n,\quad\,\,
\ZX_1 = \sum_{n=1}^7 \beta_n \mbb{s}_n,\nonumber \\
\QT_{1}^\mu &= \sum_{n=1}^5 \theta_n \mbs{v}_n^\mu,\! \quad 
\RT_1^\mu = \sum_{n=1}^5 \varrho_n \mbs{v}_n^\mu, \quad\,\,
\JJ_{1}^\mu =\sum_{n=1}^5 \gamma_n \mbs{v}_n^\mu, \quad\,
\XX_1^\mu = \sum_{n=1}^5 \vsi_n \mbs{v}_n^\mu, \label{full_set}\\
\TT_{1}^{\mu\nu} &= - \eta \sigma^{\mu\nu}_u\, .\nonumber
\end{align}
The collection of seventy parameters $\varepsilon_n, \pi_n, \varphi_n, \ldots$ provides an exhaustive set of transport parameters for the conformal superfluid at first order in the hydrodynamic derivative expansion. The transport parameters are state dependent i.e.\ are functions of $T, \m$ and $\z$. Note that the transport parameters appear in a very specific combination for $\mc{V}_1$ to ensure the tracelessness of the energy-momentum tensor. We also note that this set of seventy transport parameters constitutes the maximal possible set - many of these can be set to zero, which we will proceed to do in the following sections.

Our focus in the present work is to understand the causality and stability properties of relativistic superfluid hydrodynamics in the absence of background sources. While enumerating the independent scalars, vectors and tensors at first order in derivatives in the discussion above, we had already set the gauge field to vanish. Furthermore, since we have now computed the correct combinations of one derivative objects which carry well defined Weyl weights, in what follows we will turn off the background metric as well and specialize to flat spacetime i.e.~$g_{\m\n} = \eta_{\m\n}$ for the subsequent discussion. In the absence of background sources, the hydrodynamic equations governing the superfluid are simply given by
\begin{subequations}
\begin{align}
\del_\m T^{\m\n} = 0\, ,\label{dyn_eq_1}\\
\del_\m J^\m = 0\, ,\label{dyn_eq_2}\\
\del_\m \xi_\nu - \del_\nu\xi_\m = 0\, , \label{eq_for_xi}
\end{align}
\label{hydro_eqs_bulk}
\end{subequations}
with covariant derivatives in one derivative quantities eqs.\ \eqref{scalar_data_conf}, \eqref{vector_data_conf} and \eqref{tens_1} simply replaced by ordinary partial derivatives, i.e.~$\N_\m \rightarrow \del_\m$.

\subsection{Frame transformations and frame invariants}
\label{frames}
We shall now address an important ambiguity that exists in the hydrodynamic description of an out of equilibrium state. The hydrodynamic variables $T, \mu, u^\m$ and $\z^\m$ are well defined objects in thermal equilibrium. However, out of equilibrium, there is no \textit{a priori} definition for these quantities, and two different choices that differ from one another at $\mc{O}(\del)$ are equally valid. Thus, one may have two different definitions of the temperature, $T$ and $T'$, for the out of equilibrium state, both of which work equally well as long as their difference $T' - T \equiv \d T \sim \mc{O}(\del)$, and therefore vanishes in the limit of thermal equilibrium. To avoid this ambiguity, one has to make a choice for the out of equilibrium definition for the hydrodynamic variables. Each such choice constitutes a \emph{hydrodynamic frame}. 

Physical quantities such as $T^{\m\n}, J^\m$ and $\xi^\m$, which have a well-defined microscopic definition, are independent of the choice of the hydrodynamic frame. However, their decomposition in terms of hydrodynamic variables and their derivatives provided by the constitutive relations eq.\ \eqref{eq:most_gen_decomp} is sensitive to the choice of frame. To understand this better, consider a frame transformation given by
\be
\label{frame_tran_1}
\begin{split}
&T \rightarrow T' = T + \d T\, ,\quad  \m \rightarrow \m' = \m + \d \m\, ,\\
&u^\m \rightarrow u'^\m = u^\m + \d u^\m = u^\m + \d u \, z^\m + \d u^\m_\perp\, ,\\
&\z^\m \rightarrow \z'^\m = \z^\m + \d\z^\m = \z^\m + \z\, \d u\, u^\m + \d\z \, z^\m + \d\z^\m_\perp\, ,
\end{split}
\ee
where the perturbations $\d T, \d\m, \ldots$ are $\mathcal{O}(\del)$ quantities. Further, $\d u^\m_\perp, \d\z^\m_\perp$ lie in the plane transverse to both $u^\m$ and $z^\m$, and the particular form taken by $\d u^\m$ and $\d\z^\m$ is as such to satisfy $u'\cdot u' = -1$ and $u' \cdot \z' = 0$, up to $\mathcal{O}(\del^2)$ terms. Since $\z^\m = \z z^\m$, we can further decompose the frame transformation of $\z^\m$ into frame transformations for $\z$ and $z^\m$,
\be
\z \rightarrow \z' = \z + \d\z\, , \quad z^\m \rightarrow z'^\m = z^\m + \d z^\m = z^\m + \d u\, u^\m + \f{1}{\z}\, \d\z^\m_\perp\, .
\label{frame_tran_2}
\ee

Consider now the constitutive relation for the energy-momentum tensor, eq.\ \eqref{g_decomp_1}. Under the change of frame given by eqs.\ \eqref{frame_tran_1} and \eqref{frame_tran_2}, the zeroth order quantities transform simply as dictated by their arguments. For instance, $\e(T,\m,\z) \rightarrow \e'(T',\m',\z') = \e(T+\d T, \m + \d\m, \z+\d\z)$, and so on. At the same time, the transformation properties of $\mc{O}(\del)$ quantities, such as $\mc{E}_1 \rightarrow \mc{E}'_1$, $\mc{P}_1 \rightarrow \mc{P}'_1$, and so on, can be determined by demanding that the energy-momentum tensor stays invariant under the change of frame. This yields
{\allowdisplaybreaks
\begin{subequations}
\label{FTTcomps}
\begin{align}
&\mc{E}'_1 = \mc{E}_1 - \f{\del \e}{\del T}\, \d T - \f{\del\e}{\del\m}\, \d\m - \f{\del\e}{\del\z} \, \d\z - 2\m\tilde{\rho}\, \d u\, ,\label{E_tran}\\
&\mc{V}'_1 = \mc{V}_1 - \f{\del \tilde{\e}}{\del T}\, \d T - \f{\del\tilde{\e}}{\del\m}\, \d\m - \f{\del\tilde{\e}}{\del\z} \, \d\z - 2\m\tilde{\rho}\, \d u\, ,\label{V_tran}\\
&\mc{P}'_1 = \mc{P}_1 - \f{\del P}{\del T}\, \d T - \f{\del P}{\del\m}\, \d\m - \f{\del P}{\del\z} \, \d\z \, ,\label{P_tran}\\
&\mc{U}'_1 = \mc{U}_1 - \m \f{\del \tilde{\rho}}{\del T}\, \d T - \left(\tilde{\rho} + \m \f{\del\tilde{\rho}}{\del\m}\right) \d\m - \m \f{\del\tilde{\rho}}{\del\z} \, \d\z - \left(\e+\tilde{\e}\right) \d u\, ,\label{U_tran}\\
&\mc{Q}'^\m_1 = \mc{Q}_1^\m - \left(\e+P\right) \d u^\m_\perp - \f{\m\tilde{\rho}}{\z} \,\d\z^\m_\perp\, ,\label{Q_tran}\\
&\mc{R}'^\m_1 = \mc{R}_1^\m - \m\tilde{\rho} \, \d u^\m_\perp + \tilde{\rho} \, \d\z^\m_\perp\, ,\label{R_tran}\\
&\mc{T}'^{\m\n}_1 = \mc{T}^{\m\n}_1\, . \label{T_tran}
\end{align}
\end{subequations}}
Similarly, the frame invariance of the current $J^\m$ in eq.\ \eqref{g_decomp_2} yields
\begin{subequations}
\label{FTJcomps}
\begin{align}
&\mc{N}'_1 = \mc{N}_1 - \f{\del \rho}{\del T}\, \d T - \f{\del\rho}{\del\m}\, \d\m - \f{\del\rho}{\del\z} \, \d\z - \tilde{\rho}\, \d u\, ,\label{N_tran}\\  
&\mc{S}'_1 = \mc{S}_1 - \f{\del\tilde{\rho}}{\del T}\, \d T - \f{\del\tilde{\rho}}{\del\m}\, \d\m - \f{\del\tilde{\rho}}{\del\z} \, \d\z - \rho\, \d u\, ,\label{S_tran}\\
&\mc{J}'^\m_1 = \mc{J}_1^\m - \rho \, \d u^\m_\perp - \f{\tilde{\rho}}{\z} \, \d\z^\m_\perp\, .\label{J_tran}
\end{align}
\end{subequations}
Finally, demanding the frame invariance of $\xi^\m$, eq.\ \eqref{g_decomp_3}, under the change of frame given by eqs.\ \eqref{frame_tran_1} and \eqref{frame_tran_2} yields
\begin{subequations}
\label{FTXcomps}
\begin{align}
&\mc{M}'_1 = \mc{M}_1 - \d\m + \z \,\d u\, ,\label{M_tran}\\
&\mc{Z}'_1 = \mc{Z}_1 - \d\z + \m \,\d u \, ,\label{Z_tran}\\
&\mc{X}'^\m_1 = \mc{X}^\m_1 + \m\, \d u^\m_\perp - \d\z^\m_\perp \, .\label{X_tran}
\end{align}
\end{subequations}

As is evident from eq.\ \eqref{T_tran}, the symmetric, transverse, traceless $\mc{O}(\del)$ tensor $\mc{T}^{\m\n}_1$ is a frame invariant object. There also exist frame invariant transverse vectors and scalars. For instance, by looking at the frame transformations in eqs.\ \eqref{Q_tran}, \eqref{R_tran}, \eqref{J_tran} and \eqref{X_tran}, it is easy to see that we can construct two frame invariant transverse vectors. A convenient choice is
\begin{subequations}\label{eq:frame-invar-vectors}
\begin{align}
\mc{L}^\m &\equiv \mc{Q}_1^\m - \mbs{e}_1 \mc{J}_1^\m - \mbs{e}_2 \mc{X}_1^\m \, ,\\
\mc{K}^\m &\equiv \mc{R}^\m_1 - \mbs{j}_1 \mc{J}^\m_1 - \mbs{j}_2 \mc{X}^\m_1\, , 
\end{align}
\end{subequations}
with
\be
\mbs{e}_1 = \f{(\e+P)\z+\m^2\tilde{\rho}}{\m \tilde{\rho} + \z \rho}\, , \quad \mbs{e}_2 = - \f{Ts\tilde{\rho}}{\m \tilde{\rho} + \z \rho}\, , \quad \mbs{j}_1 = 0\, , \quad \mbs{j}_2 = - \tilde{\rho}\, ,
\ee
where we have made use of the thermodynamic relation $\e+P = Ts + \m\rho$, eq.\ \eqref{etedefs}. Similarly, we can construct four frame invariant $\mc{O}(\del)$ scalars. These can be chosen to be
\be
\begin{split}\label{eq:frame-invar-scalars}
\mc{F} &\equiv \mc{V}_1 - \mbs{a}_1 \mc{E}_1 - \mbs{a}_2 \mc{N}_1 - \mbs{a}_3 \mc{S}_1 - \mbs{a}_4 \mc{Z}_1\, , \quad \mc{G} \equiv \mc{P}_1 - \mbs{b}_1 \mc{E}_1 - \mbs{b}_2 \mc{N}_1 - \mbs{b}_3 \mc{S}_1 - \mbs{b}_4 \mc{Z}_1\, ,\\
\mc{L} &\equiv \mc{U}_1 - \mbs{c}_1 \mc{E}_1 - \mbs{c}_2 \mc{N}_1 - \mbs{c}_3 \mc{S}_1 - \mbs{c}_4 \mc{Z}_1\, , \quad \mc{H} \equiv \mc{M}_1 - \mbs{d}_1 \mc{E}_1 - \mbs{d}_2 \mc{N}_1 - \mbs{d}_3 \mc{S}_1 - \mbs{d}_4 \mc{Z}_1\, .
\end{split}
\ee
The quantities $\mbs{a}_i, \mbs{b}_i, \mbs{c}_i, \mbs{d}_i$, $i = 1, \ldots 4$ are thermodynamic objects, which can be straightforwardly obtained using the known frame transformations in eqs.\ \eqref{FTTcomps}, \eqref{FTJcomps} and \eqref{FTXcomps} computed above, and demanding the frame invariance of $\mc{F}, \mc{G}, \mc{L}$ and $\mc{H}$. The resulting expressions are quite large, and we omit presenting them here for the sake of brevity. 

In the conformal limit, the tracelessness of the energy-momentum tensor leads to the relation $\mc{E}_1 - \mc{V}_1 = (d-1) \mc{P}_1$, implying that $\mc{F} = -(d-1) \mc{G}$,\footnote{
This may be seen straightforwardly by considering the definitions of ${\cal F}$ and ${\cal G}$ in a frame with
\be
\mc{E}_1 = \mc{N}_1 = \mc{S}_1 = \mc{Z}_1 = \mc{J}_1^\mu = \mc{X}_1^\mu = 0\, ,
\ee
which we will presently introduce as the ``LLCP frame" in section \ref{LLCPframe}. In this frame, ${\cal F} = {\cal V}_1^{\rm LLCP}$, ${\cal G} = {\cal P}_1^{\rm LLCP}$, and the relationship $\mc{E}_1 - \mc{V}_1 = (d-1) \mc{P}_1$ becomes $-{\cal V}_1^{\rm LLCP} = \lr{d-1} {\cal P}_1^{\rm LLCP}$. This implies that ${\cal F} = {\cal V}_1^{\rm LLCP} = - \lr{d-1} {\cal P}_1^{\rm LLCP} = - \lr{d-1} {\cal G}$, and hence ${\cal F} = - \lr{d-1} {\cal G}$. However, ${\cal F}$ and ${\cal G}$ are both frame invariant quantities, thereby implying that this relation must be true in any hydrodynamic frame.
} and thus leads to one less frame invariant scalar at $\mc{O}(\del)$.

\subsection{The LLCP frame}
\label{LLCPframe}
The Landau-Lifshitz-Clark-Putterman (LLCP) frame was introduced in \cite{Landau-Lifshitz, ClarkThesis, putterman1974} for understanding superfluid hydrodynamics, and was more recently investigated in \cite{Bhattacharya:2011eea} to construct the fluid/gravity map for dissipative superfluid hydrodynamics at the first order in the derivative expansion. In the present section, we will look at the defining features of the LLCP frame. Subsequently, in section \eqref{LLCP_problems}, we will argue that this frame is acausal, and therefore also unstable, motivating one to use the BDNK procedure of working in a general frame for superfluids as well.

Concretely, in the notation of the previous section, the LLCP frame corresponds to the choice
\be
\mc{E}_1 = \mc{N}_1 = \mc{S}_1 = \mc{Z}_1 = \mc{J}_1^\mu = \mc{X}_1^\mu = 0
\label{LLCP_choice}
\ee
in the most general constitutive relations eq.\ \eqref{eq:most_gen_decomp}. Since we are interested in conformal superfluids, we further have $\mc{V}_1 = -(d-1) \mc{P}_1$ in the constitutive relation eq.\ \eqref{g_decomp_1}. One is therefore left with three independent scalar corrections - $\mc{P}_1, \mc{U}_1, \mc{M}_1$; two vector corrections - $\mc{Q}_1^\m, \mc{R}_1^\m$; and one tensor correction - $\mc{T}_1^{\m\n}$, at first order in derivatives in the constitutive relations eq.\ \eqref{eq:most_gen_decomp} in the LLCP frame. The three scalar corrections can be expanded in the basis of the seven Weyl covariant scalars eq.\ \eqref{scalar_data_conf}, as in eq.\ \eqref{full_set}. Similarly, the two vector corrections can be expanded in the Weyl covariant basis of transverse vectors given in eq.\ \eqref{vector_data_conf}, as done in eq.\ \eqref{full_set}. Including the tensor contribution, this leads to a set of 32 transport parameters at the first order in the derivative expansion for the conformal superfluid in the LLCP frame. 

Operationally, one can go onshell and reduce the number of independent transport parameters by utilizing relations between the one derivative scalars and vectors in eqs.\ \eqref{scalar_data_conf} and \eqref{vector_data_conf} obtained by imposing the hydrodynamic equations on the ideal superfluid constitutive relations eq.\ \eqref{ideal_currs_bulk}. We get the following relations amongst the scalars, 
\begin{subequations}
\label{eq:scalar_rels}
\begin{align}
u_\n \del_\m T^{\m\n}_{\rm ideal} = 0 &\Rightarrow (d+1) \e 
\, \mbb{s}_1 + T \f{\del\e}{\del\m} \, \mbb{s}_2 + T \f{\del\e}{\del\z} \, \mbb{s}_3 + T \tilde{\rho} \, \mbb{s}_4 + 2 \m\tilde{\rho}\, \mbb{s}_5 \nonumber \\
&\hspace{49mm} - \z \tilde{\rho} \, \mbb{s}_6 + \m T^d \, \mbb{s}_7 = 0, \\
z_\n \del_\m T^{\m\n}_{\rm ideal} = 0 &\Rightarrow (d+1) \m\tilde{\rho} \, \mbb{s}_1 + T\left(\tilde{\rho} + \m \f{\del\tilde{\rho}}{\del\m}\right) \mbb{s}_2 + \m T \f{\del\tilde{\rho}}{\del\zeta} \, \mbb{s}_3 \nonumber \\
&\qquad\,\, + T\rho \, \mbb{s}_4 + (\e+\tilde{\e}) \, \mbb{s}_5 + \m\tilde{\rho} \, \mbb{s}_6 - \z T^d \, \mbb{s}_7 = 0,\\
\del_\m J^\m_{\rm ideal} = 0 &\Rightarrow d\rho \, \mbb{s}_1 + T \f{\del\rho}{\del\m}\, \mbb{s}_2 + T \f{\del\rho}{\del\z} \,\mbb{s}_3 + \tilde{\rho} \, \mbb{s}_5 + T^d \, \mbb{s}_7 = 0, \\
u^\m z^\n \left(\del_\m \xi_\n^{\rm id.} - \del_\n \xi_\m^{\rm id.} \right) = 0 &\Rightarrow \z\, \mbb{s}_1 + T \, \mbb{s}_3 - T\, \mbb{s}_4 - \m \, \mbb{s}_5 + \z \, \mbb{s}_6 = 0.
\end{align}
\end{subequations}
Similarly, we have the transverse vector relations
\begin{subequations}
\label{eq:vector_rels}
\begin{align}
\tilde{\D}^\a_\n \del_\m T^{\m\n}_{\rm ideal} = 0 &\Rightarrow (\e+P) \mbs{v}_1^\m + T\rho\, \mbs{v}_2^\m + T \tilde{\rho} \, \mbs{v}_5^\m = 0,\\
\tilde{\D}^{\a\m} u^\n \left(\del_\m \xi_\n^{\rm id.} - \del_\n \xi_\m^{\rm id.}\right) = 0 &\Rightarrow \m \, \mbs{v}_1^\m + T \, \mbs{v}_2^\m - \z \, \mbs{v}_4^\m  = 0,\\
\tilde{\D}^{\a\m} z^\n \left(\del_\m \xi_\n^{\rm id.} - \del_\n \xi_\m^{\rm id.}\right) = 0 &\Rightarrow \m \, \mbs{v}^\m_4 - T \, \mbs{v}_5^\m  = 0. 
\end{align} 
\end{subequations}
Employing the relations above, the number of independent Weyl covariant scalars at first order in derivatives comes down to three, whereas the number of independent Weyl covariant one derivative transverse vectors comes down to two. Different choices for which three scalars and two vectors are kept in writing the constitutive relations leads to different members within the LLCP family of frames.\footnote{Note that the background equations \eqref{eq:vector_rels} do not involve $\mbs{v}_3^\m$. It must therefore be chosen as one of the two independent transverse vectors.} Though all these choices are physically equivalent, they are mathematically distinct at the level of the partial differential equations they give rise to when one writes the hydrodynamic equations. A possible convenient choice, also utilized in \cite{Bhattacharya:2011eea}, is to work with the scalars $(\mbb{s}_5, \mbb{s}_6, \mbb{s}_7)$ and the vectors $(\mbs{v}_1, \mbs{v}_3)$, in terms of which the LLCP frame constitutive relations become
\begin{align}
\PT_1 &= \sum_{n=5}^7 \bar{\pi}_n \mbb{s}_n, \quad \VT_1= - \lr{d-1}\sum_{n=5}^7 \bar{\pi}_n \mbb{s}_n, \quad \UT_1 = \sum_{n=5}^7 \bar{\varphi}_n \mbb{s}_n, \quad {\cal M}_1 = \sum_{n=5}^7 \bar{\alpha}_n \mbb{s}_n, \nonumber \\
\QT_{1}^\mu &= \sum_{n=1, 3} \bar{\theta}_n \mbs{v}_n^\mu,\! \quad 
\RT_1^\mu = \sum_{n=1, 3} \bar{\varrho}_n \mbs{v}_n^\mu,  \qquad \TT_{1}^{\mu\nu} = - \eta \sigma^{\mu\nu}_u\, . \label{full_set_LLCP}
\end{align}
The overhead bars on the transport parameters denote the LLCP frame. At this stage, there appear to be 14 different transport parameters for the conformal superfluid at the first order. The Onsager reciprocity condition, which follows by demanding the time-reversal invariance of the underlying microscopic theory, leads to a further reduction in their number. As discussed in detail in appendix \ref{entropy_LLCP}, the reciprocity condition leads to the relations
\be
\label{Onsager}
d\bar{\pi}_5 = - \bar{\vp}_6\, , \quad d\bar{\pi}_7 = -T^d \bar{\a}_6\, , \quad \bar{\vp}_7 = T^d \bar{\a}_5\, ,  \quad \bar{\th}_3 = \bar{\varrho}_1.
\ee
We can therefore eliminate $\bar{\vp}_6, \bar{\a}_5, \bar{\a}_6$ and $\bar{\varrho}_1$ in favour of the other four transport parameters. Thus, we are left with only 10 independent transport parameters for the conformal superfluid at the first order. These are further constrained by the requirement that the dissipative effects they generate should lead to a non-negative production of entropy onshell. This gives the constraints
\begin{align}
&\bar{\vp}_5 \le 0\, , \quad \bar{\pi}_6 \ge 0\, , \quad \bar{\a}_7 \le 0\, , \quad \bar{\pi}_5^2 \le -\frac{1}{d} \bar{\vp}_5 \bar{\pi}_6\, , \quad \bar{\pi}_7^{\,2} \le - \frac{T^d}{d} \bar{\pi}_6 \bar{\a}_7\, , \quad \bar{\vp}_7^{\,2} \le T^d \bar{\vp}_5 \bar{\a}_7 \, , \nonumber \\
&\bar{\th}_1 \le 0\, , \quad \bar{\varrho}_3 \le 0\, , \quad \bar{\th}_3^{\,2} \le \bar{\th}_1 \bar{\varrho}_3 \, , \quad \eta \ge 0\, , \label{2ndLawConstraints}
\end{align}
as discussed in detail in appendix \ref{entropy_LLCP}. We refer to these transport parameters that appear in the entropy current as ``physical transport coefficients."\footnote{The physical transport coefficients are differentiated from transport parameters by the fact that one can write down Kubo formulae for the physical transport coefficients~\cite{Kovtun:2012rj}, and thereby extract their values from an underlying microscopic theory. No such process can be undertaken for a general transport parameter: they are ``purely frame", i.e. they have no microscopic definition, and act as UV regulators to imbibe the hydrodynamics with better stability and causality properties. It is possible for physical transport coefficients to not appear in the entropy current, but still have Kubo formulae. Examples of such non-hydrostatic, non-dissipative transport coefficients are Hall conductivities and viscosities in relativistic magnetohydrodynamics~\cite{Hernandez:2017mch}.}


\section{LLCP frame and acausality}
\label{sec:basics}
Having overviewed the groundwork for understanding first order dissipative relativistic superfluid hydrodynamics in the previous section, we now take a closer look at the out of equilibrium dynamics, with a particular focus on the LLCP frame. In section \ref{sec:dispersion}, we begin by examining the dispersion relations for the linearized system of superfluid hydrodynamic equations about a homogeneous equilibrium state. In section \ref{stable_causal}, we collect the main results from the existing literature and recapitulate the necessary stability and causality conditions that should be met by the dispersion relations and the associated hydrodynamic modes. Then, in section \ref{LLCP_problems}, we argue that out of equilibrium dynamics in the LLCP frame violates these conditions, necessitating the use of a more general hydrodynamic frame to meet the stability and causality requirements, in the spirit of the BDNK program \cite{Bemfica:2017wps, Kovtun:2019hdm, Bemfica:2019knx, Hoult:2020eho}. 

\subsection{Dispersion relations}
\label{sec:dispersion}
\subsubsection{Linear perturbations about homogeneous equilibrium}
The equations of motion governing superfluid hydrodynamics in the absence of sources, \eqref{hydro_eqs_bulk}, admit time-independent homogeneous solutions. One may investigate the linear response of the system by considering plane-wave perturbations about this homogeneous equilibrium state, which have the general form
\[
\mathbb{U}(x) = \mathbb{U}_0 + \delta \mathbb{U}(\omega,\kv) \, e^{-i \omega t + i \kv{\cdot}\xv} ,
\]
where $\mathbb{U}(x) \equiv \{T, \mu, u^\mu, \zeta^\mu\}$ is a column vector comprising the hydrodynamic degrees of freedom, with $\mathbb{U}_0$ denoting their values in the homogeneous equilibrium state while $\d\mathbb{U}$ denotes perturbations away from equilibrium, $\kv$ is the wave-vector, and $\omega$ is the angular frequency. One finds that these plane-waves only constitute solutions to the linearized equations of motion if the $\omega$ are particular functions of $\kv$. These ``dispersion relations", or ``modes", $\omega_{(n)} = \omega_{(n)}(\kv)$,\footnote{The subscript $n$ runs over the different modes.} contain important information about the system, including its behaviour in the hydrodynamic regime. To compute the dispersion relations, one demands that the linearized equations of motion be satisfied for non-zero perturbations about the homogeneous equilibrium state i.e.~when $\d {\mathbb U}(\o, {\bf k}) \neq 0$, generically leading to a set of equations of the form
\[
{\mathbb M}_{\rm AB} \, \delta \mathbb{U}^{\rm B} e^{-i \omega t + i \kv{\cdot}\xv} = 0\, ,
\]
where ${\rm A,B}$ are matrix indices, and the elements of the matrix $\mathbb{M}$ too are functions of $(\o, {\bf k})$. The dispersion relations then correspond to the solutions of the equation
\begin{equation}\label{spectral_curve}
\det\lr{{\mathbb M}} \equiv F(\omega, \kv, \mathbb{U}_0) = 0\, ,
\end{equation}
where the function $F$ is referred to as the ``spectral curve" of the system.\footnote{The name is chosen to reflect the fact that $F$ can be thought of as an analytic curve in $(\omega, \kv)$, and also to disambiguate with uses of the phrase ``spectral function" in the literature.} However, this is an over-simplification, as one finds that the components of $\delta \mathbb{U}$ are not all independent. The reason for this is that, in addition to the $d$ dynamical equations in \eqref{eq_for_xi}, there are also $d\times\lr{d-1}/2$ constraint equations. There are also, of course, the algebraic constraints $u{\cdot}u = -1$ and $u{\cdot}z = 0$ that the components of $\mathbb{U}$ must satisfy.

To perform the linear analysis, we consider the following time-independent homogeneous solution to the equations of motion:
\begin{equation}
T = T_0, \quad \mu = \mu_0,\quad  u^\mu_0 = \delta_t^\mu, \quad \zeta^\mu_0 = \zeta_0 \,\delta_z^\mu\, ,
\label{homo_state}
\end{equation}
where $T_0, \m_0$ and $\z_0$ are constants. We refer to this as the \emph{rest frame}.\footnote{This is not to be confused with the notion of hydrodynamic frames discussed in section \ref{sec:constitutive}.} Of course, the choice to align $\zeta^\mu$ with the $z$-axis is completely arbitrary and can always be carried out using the rotational symmetry of the setup. 

Let us now impose the constraints from eq.~\eqref{eq_for_xi} on the perturbations $\delta \mathbb{U}(\o, \kv)$ about the rest frame. To begin with, let us consider the constraints that follow by truncating the constitutive relation for $\xi^\m$ at zeroth order in derivatives. One finds that the constraint equations in \eqref{eq_for_xi}, together with $u{\cdot}u = -1$ and $u{\cdot}\zeta = 0$, mean that we can decompose the perturbations to the fluid velocity $u^\mu$ and transverse superfluid velocity $\zeta^\mu$ as
\begin{equation}
\begin{split}
\delta u^\mu(\omega, \kv) &= \lr{z_0{\cdot} \delta u} \delta_z^\mu + \tD^{\mu\nu}\delta u_\nu,\\
\delta \zeta^\mu(\omega, \kv) &= \lr{\zeta_0{\cdot}\delta u} \delta_t^\mu + \mu_0 \tD^{\mu\nu}\delta u_\nu + \delta \hat{\zeta} \hat{k}^\mu,
\end{split}
\label{main_constraints}
\end{equation}
where $\hat{k}^\mu = \delta^\mu_i k^i/|\kv|$ points in the direction of the wave-vector. As such, there are only three independent components of $\delta u^\mu$ and only one independent component of $\delta \zeta_\mu$ in the rest frame.\footnote{This is reflective of the fact that the underlying degree of freedom is actually the (scalar) Goldstone.}

Now, let us return to first order in the derivative expansion for the constitutive relation of $\xi^\m$, and work out the constraints. The algebraic constraints $u{\cdot}u = -1$ and $u{\cdot}\zeta = 0$ are unaffected. However, the derivative constraints in \eqref{eq_for_xi} are now, in general, frame dependent, and contain both first order time derivatives and second order spatial derivatives. The components of $\delta \mathbb{U}(\omega,\kv)$ are therefore related to one another in an $\omega$-, $|\kv|$-dependent fashion, unlike at zeroth order where the relations depended solely on $\hat{k}$. While this is not \textit{a-priori} a problem, it does make the analysis significantly more challenging. It is therefore convenient to choose a family of hydrodynamic frames in which the solutions to the constraint equations \eqref{eq_for_xi} with $\xi^\m$ incorporating first order derivative corrections are the same as the ones when $\xi^\mu$ was truncated to zeroth order i.e.~eq.~\eqref{main_constraints}. This is the case if
\begin{equation}
\label{eq:constraint_frames}
\begin{split}
\beta_1 &= \frac{\zeta_0}{T_0} \beta_3, \quad \beta_2 = 0, \quad \beta_4 = \varsigma_2,\quad \beta_5 = - \frac{\mu_0}{T_0} \beta_3, \quad  \beta_6 = \frac{\zeta_0}{T_0} \beta_3, \quad     \beta_7 = 0\,,\\
\varsigma_1 &= - \frac{\mu_0}{T_0}\beta_3, \quad \varsigma_3 = 0, \quad  \varsigma_4 = \lr{\frac{\zeta_0}{T_0}} \beta_3 - \lr{\frac{\mu_0}{T_0}} \varsigma_5\,.
\end{split}
\end{equation}
There are therefore only three independent transport parameters in ${\cal Z}$, ${\cal X}_\mu$ if we would like the constraint equations to have the simple solution eq.~\eqref{main_constraints}. As is clear from above, we have taken these independent parameters to be $\beta_3$, $\varsigma_2$, and $\varsigma_5$. For lack of better terminology, we refer to this class of hydrodynamic frames as the \emph{constraint frames}. The LLCP frame is automatically a member of the constraint frames, as ${\cal Z}_1 = 0$ and ${\cal X}_1^\mu = 0$.

\subsection{Overview of the stability and causality requirements}
\label{stable_causal}
In general, there are two dynamical properties that we ought to demand of any relativistic hydrodynamic theory. Firstly, we should demand that plane-wave perturbations about homogeneous equilibria decay with time. Secondly, we should demand that wavefronts in the theory travel no faster than the speed of light. In other words, we should demand that the theory be linearly stable, and causal. These properties can be identified at the level of the dispersion relations for the system.\footnote{See \cite{Gouteraux:2024adm} for an interesting discussion where the authors argue that stability of the thermal equilibrium state vis-\`{a}-vis positivity of the static susceptibility matrix, along with the local second law of thermodynamics ensuring positivity of entropy production, are conditions sufficient to ensure that the physical hydrodynamic modes lie in the lower half of the complex frequency plane and are stable, eq.~\eqref{stab_condition}. However, depending upon the frame, the dispersion relations also include non-hydrodynamic modes, and to arrive at a UV-complete formulation one should render these modes stable too by a judicious choice of frame, which is the underlying philosophy of the BDNK program.}\\

\noindent $\bullet$ \emph{\underline{Linear stability}:} In order for a mode to be stable, we must demand that for $\kv \in \mathbb{R}^3$,
\begin{equation}\label{stab_condition}
{\rm Im} (\omega) \leq 0.
\end{equation}
In particular, since $\omega$ is a function of $\kv$, this must hold for \textit{any} value of $\kv$.\\

\noindent $\bullet$ \emph{\underline{Causality}:} In order for the system to be (linearly) causal, we must demand that the following three criteria hold on the dispersion relations \cite{Hoult:2023clg} for $\kv \in \mathbb{R}^3$:
\begin{equation}\label{large_k_caus_constraints}
\begin{gathered}
0 \leq \lim_{|\kv| \to \infty} \frac{|{\rm Re}\lr{\omega(\kv)}|}{|\kv|} < 1, \qquad\lim_{|\kv| \to \infty} \frac{{\rm Im} \lr{\omega(\kv)}}{|\kv|} = 0\,, \\
{\cal O}_\omega(F(\omega,\kv, \mathbb{U}_0)) = {\cal O}_{|\kv|} \lr{F(\omega = \mathfrak{a} |\kv|, k_j = \mathbf{s}_j |\kv|,\mathbb{U}_0)}\,,
\end{gathered}
\end{equation}
where ${\cal O}_\xi$ indicates the order of the polynomial in $\xi$, $\mathfrak{a}$ is an arbitrary non-zero constant, and $\mathbf{s}_j$ is a unit vector.  The first condition ensures that modes stay within the lightcone at large $|\kv|$; the second and third conditions ensure the (at least weak) hyperbolicity of the system. In particular, the third condition serves to eliminate any infinite speed ``stray modes" in the system which are not visible in the rest frame.\footnote{These stray modes are the reason that the Landau frame in uncharged hydrodynamics may look unproblematic at first glance when studying linear perturbations about the rest frame. They propagate at infinite speed, and don't show up in the linearized analysis when solving just for $\omega = \omega(\kv)$ in the rest frame.} In particular, as we will discuss later, such stray modes do exist in the LLCP frame.

It turns out that if both eqs.~\eqref{stab_condition} and \eqref{large_k_caus_constraints} hold for linear perturbations about the rest frame, they will continue to hold in a boosted reference frame as well  \cite{Gavassino:2021owo,Gavassino:2023myj}. In other words, showing stability in the rest frame for a causal system is sufficient to ensure stability of the system in any other reference frame. This significantly reduces the amount of work required to show the stability and causality of any given system in general.

Finally, for completeness, we note that the stability condition and the first two causality conditions may be combined into a single algebraic condition on the dispersion relations for $\omega, |\kv| \in \mathbb{C}$ \cite{Heller:2022ejw,Gavassino:2023myj,Heller:2023jtd}, given by
\begin{equation}\label{covariant_stability}
    {\rm Im}(\omega) \leq | {\rm Im}(\kv) |.
\end{equation}
Demanding this constraint will enforce certain conditions on the form of the dispersion relations, in both the small-$|\kv|$ \cite{Heller:2022ejw,Gavassino:2023myj,Heller:2023jtd} and large-$|\kv|$ limits \cite{Hoult:2023clg,Wang:2023csj}. We will not be making direct use of this condition in the present work; however, its use in constraining hydrodynamic theories is an interesting open area.

With these criteria in mind, we now argue that the LLCP frame is acausal, and therefore unstable in a boosted reference frame.

\subsection{Acausality and instability of the LLCP frame}
\label{LLCP_problems}
Let us perform a linearized analysis of perturbations about the homogeneous equilibrium state, eq.~\eqref{homo_state}, in the LLCP frame. We begin by looking at the roots of the spectral curve $F(\omega,\kv,\mathbb{U}_0)$ of the system in the limit $|\kv| \to 0$. We have
\begin{equation}
    \lim_{|\kv| \to 0}F_{\rm LLCP}(\omega,\kv,\mathbb{U}_0) =  \omega^6 \lr{\theta_1 \zeta_0 \omega + i \lr{\lr{\epsilon+P} \zeta_0 + \mu_0^2 \tilde{\rho}}}^2 \lr{i a \omega + b} = 0\, ,
\end{equation}
where $a = \lr{\pder{\epsilon}{T} \pder{\rho}{\mu} - \pder{\epsilon}{\mu} \pder{\rho}{T}} \varphi_5$, and $b$ is a purely thermodynamic function. We see from above that the system has three gapped (``non-hydrodynamic") modes, i.e.~satisfying $\lim_{|\kv|\rightarrow 0} \omega(\kv) \neq 0$, and six gapless (``hydrodynamic") modes, for which $\lim_{|\kv|\rightarrow 0} \omega(\kv) = 0$. Indeed, the spectral curve in the LLCP frame for non-zero $|\kv|$ is an order-$9$ polynomial in $\omega$. We therefore have ${\cal O}_\omega\lr{F(\omega,\kv,\mathbb{U}_0)} = 9$. However, the spectral curve, without imposing $|\kv| \to 0$, also has ${\cal O}_{|\kv|}\lr{F(\omega = \mathfrak{a} |\kv|, k_j = \mathbf{s}_j |\kv|,\mathbb{U}_0)} = 12$. The third condition in the requirements for the system to be causal, eq.~\eqref{large_k_caus_constraints}, is therefore not satisfied in the LLCP frame.

It turns out that the LLCP frame also violates the other two conditions required for causality. In the limit $|\kv| \to \infty$, there are are only five modes that go linearly in $|\kv|$. Of the remaining four modes, one behaves at $|\kv|\to\infty$ as $\omega \sim |\kv|^3$, while the remaining three take the form $\omega \sim |\kv|^{4/3}$. 
These violate the first two conditions of eq.~\eqref{large_k_caus_constraints}. The LLCP frame is thus acausal, and therefore there also exists a boosted reference frame in which it is linearly unstable \cite{Gavassino:2021owo,Gavassino:2023myj}.\footnote{In appendix \ref{mod_Landau_frame}, we take a close look at another important frame, the ``superfluid Landau frame'' of \cite{Arean:2023nnn}, which we call the \emph{modified Landau frame}, and argue that it too is acausal and unstable.} These observations motivate one to look for hydrodynamic frames where the out of equilibrium dynamics of superfluids remains stable and causal, a subject which we turn next to.


\section{Linearly stable and causal superfluid hydrodynamics}
\label{sec:new_material}
Following the BDNK procedure \cite{Bemfica:2017wps, Kovtun:2019hdm, Bemfica:2019knx, Hoult:2020eho}, we now move on to discuss conformal superfluid hydrodynamics in a general frame, demanding the linear stability and causality criteria laid out in section \ref{stable_causal} to be satisfied. 
We find a set of hydrodynamic frames that meet these criteria in certain restricted regimes, providing some of the main results of this paper. At the same time, we also highlight that some of the transport parameters that are necessary to ensure stability and causality are forced to vanish by the choice of the LLCP frame, leading to the acausality and instability issues associated with it, as observed in section \ref{LLCP_problems}. For convenience, we look for stable and causal frames within the family of constraint frames, eq.~\eqref{eq:constraint_frames}, and demand that linearized perturbations be of the form eq.~\eqref{main_constraints}.

\subsection{Linear stability criteria in a general frame}
\label{gen_stability}
Finding a frame analytically where eq.~\eqref{stab_condition} is satisfied for arbitrary $\kv$ is a highly non-trivial task. In particular, one must find a frame in which the spectral curve satisfies the so-called Routh-Hurwitz criterion at arbitrary $\kv$, where for the full BDNK system, the spectral curve is an order-12 polynomial in $\omega$ with $\kv$-dependent coefficients. The Routh-Hurwitz criterion is a condition on polynomials which, if satisfied, guarantees that all the roots of the polynomial reside in the left half of the complex-plane. Adapting the criterion to ensure that the roots lie in the lower half of the complex-plane is simple. For more information, the interested reader may refer to Appendix \ref{app:Routh-Hurwitz}. Given the analytic complexity of the problem, we leave an exhaustive search for arbitrary $\kv$ to further numerical work. In particular, the search should simplify with a particular equation of state in mind.

While the full problem may be analytically intractable, one may instead consider a simpler setting. As discussed earlier, in the limit that $|\kv| \to 0$, the dispersion relations may be classified into two categories -- the ``gapless" modes for which $\omega \to 0$ as $|\kv| \to 0$, and the ``gapped" modes for which $\omega$ remains non-zero as $|\kv| \to 0$. The spectral curve therefore factors when $|\kv| \to 0$. For the conformal superfluid, we find
\begin{equation}
\begin{split}\label{eq:gaps}
\lim_{|\kv| \to 0}F(\omega, \kv, \mathbb{U}_0) &= \omega^6 \times  {\cal G}(\omega, \mathbb{U}_0) \times  \lr{i + \frac{\beta_3}{T_0}\omega}\\ 
&\quad\times\left[ i \zeta_0 \lr{4 P - \zeta_0 \tilde{\rho}_0} + \lr{\zeta_0 \theta_1 + \mu_0 \lr{\theta_4 + \frac{\mu_0}{T_0} \theta_5}}\omega\right]^2 = 0\, ,
\end{split}
\end{equation}
where ${\cal G}(\omega, \mathbb{U}_0)$ is a rather complicated cubic polynomial which is highly dependent on the equation of state. It is of the generic form
\begin{equation}
\label{G_generic_1}
{\cal G}(\omega, \mathbb{U}_0) = \mbs{a} \lr{- i \omega}^3 + \mbs{b} \lr{- i \omega}^2 + \mbs{c} \lr{-i \omega} + \mbs{d}\, ,
\end{equation}
where $\mbs{a}$, $\mbs{b}$, $\mbs{c}$, $\mbs{d}$ are functions of the transport parameters and the thermodynamic variables. In particular, $\mbs{d}$ depends solely on the thermodynamic variables. Let us assume the equation of state is such that $\mbs{d} \neq 0$; we then divide through by $\mbs{d}$ to get
\[
{\cal G}'(\omega, \mathbb{U}_0)  = {\cal G}(\omega, \mathbb{U}_0)/\mbs{d} = \mbs{a}' \lr{-i \omega}^3 + \mbs{b}' \lr{- i \omega}^2 + \mbs{c}' \lr{- i \omega} + 1.
\]

Now, in order to ensure that the gapped solutions in eq.~\eqref{eq:gaps} are stable in the limit $|\kv| \to 0$, we must demand that ${\rm Im}\lr{\omega} < 0$. Therefore, we must have that
\begin{subequations}
\label{lin_stab_1}
\begin{align}
\beta_3 &> 0\,,\\
 \lr{4 P - \zeta_0 \tilde{\rho}_0}/ \lr{ \theta_1 + \frac{\mu_0}{\zeta_0} \lr{\theta_4 + \frac{\mu_0}{T_0} \theta_5}} &> 0\, ,\\
 \mbs{a}' > 0, \quad \mbs{b}'>0, \quad \mbs{b}' \mbs{c}' - \mbs{a}'&> 0\, .
\end{align}
\end{subequations}
The final set of inequalities come from applying the Routh-Hurwitz criterion to ${\cal G}'(\omega, \mathbb{U}_0)$. We note that the condition $\beta_3 > 0$ is violated by the LLCP frame, thereby leading to problems with stability.

These constraints may be simplified slightly if we restrict ourselves to a smaller set of hydrodynamic frames, which we refer to as the ``decoupled frames", so called due to their effect (to be described later on) at large $k$. Consider a class of frames\footnote{We note at this point that the choice of the ``decoupled frames" here has been imposed \textit{on top} of the previous choice of the constraint frames~\eqref{eq:constraint_frames}. The choices are in principle independent.} with
\begin{equation}\label{decoupled_frames}
\ve_{2,3,4,7} = \pi_{2,3,4,7} = \varphi_{2,3,4,7} = \theta_{2,4,5} = \varrho_{2,4,5} = 0, \quad \alpha_2 = -\varsigma_2, \quad \alpha_4 = - \frac{1}{T_0^2}\pder{\tilde{\rho}}{\mu} \alpha_7.
\end{equation}
Then, the linear stability requirements eq.~\eqref{lin_stab_1} become
\begin{subequations}
\label{lin_stab_2}
\begin{align}
\beta_3 &> 0,\\
\lr{4 P - \zeta_0 \tilde{\rho}}/\theta_1 &> 0,\\
\frac{1}{\Omega} \nu_2 \lr{\ve_1 \varphi_5 - \ve_5 \varphi_1} &> 0,\\
\mbs{b}' &>0,\\
\mbs{b}' \mbs{c}' - \frac{1}{\Omega} \nu_2 \lr{\ve_1 \varphi_5 - \ve_5 \varphi_1} &> 0,
\end{align}
\end{subequations}
where $\Omega$ is a purely thermodynamic factor which equals $\mbs{d}/ T_0^2$. The decoupled frames can thus provide a set of hydrodynamic frames where superfluid hydrodynamics is linearly stable in the limit $|\kv| \to 0$. 

\subsection{Linear causality criteria in a general frame}
\label{gen_causality}
Causality in a system which is governed by a set of partial differential equations is generally best understood via the theory of partial differential equations, rather than any sort of linearized perturbations about an equilibrium state. However, the presence of the constraint equations in eq.~\eqref{eq_for_xi} complicates such a non-linear analysis. As such, we will consider the causality of linearized fluctuations about an equilibrium state, applying the constraints in eq.~\eqref{eq_for_xi} directly to the perturbations in Fourier space.

To begin, it is clear from the constraints eq.~\eqref{large_k_caus_constraints} that in the large-$|\kv|$ limit $\omega$ must take the form
\[
\omega = v \,|\kv| + {\cal O}(|\kv|^{n<1})\, ,
\]
where $v = v(\hat{k}^i) \in \mathbb{R}$ is the asymptotic phase velocity and $0 \leq v^2 < 1$, with $\hat{k}^i \equiv k^i/|\kv|$ denoting components of the unit vector along $\kv$. If we look at the large-$|\kv|$ limit of the spectral curve, we find that the various linear coefficients $v_{(n)}$ are generically the roots of an order-10 polynomial and an order-2 polynomial. While the order-2 polynomial is simple to deal with, calculations with the order-10 polynomial are analytically intractable. 

The situation simplifies significantly, however, if we use the decoupled frames defined in eq.~\eqref{decoupled_frames}. Then the controlling equations for the linear coefficient $v$ take the form 
\begin{subequations}
\label{caus_crit_1}
\begin{align}
    \theta_1 v^2 - \cos(\theta) \lr{\theta_3 + \varrho_1} v + \varrho_3 \cos^2(\theta) - \eta \sin^2(\theta) &= 0, \label{shear_mode}\\
    \beta_3 v^2 - \alpha_3 \cos(\theta) v - \frac{\alpha_7}{2T_0^2 \zeta_0} \lr{\pder{\tilde{\rho}}{\zeta} \zeta_0 + \lr{\pder{\tilde{\rho}}{\zeta} \zeta_0 - 2 \tilde{\rho}} \lr{\cos^2(\theta) - \sin^2(\theta)}} &= 0, \label{charge_1}\\
    \nu_2 v^2 - \lr{\lambda_2 + \nu_4 + \frac{1}{T_0^2} \pder{\tilde{\rho}}{\mu} \nu_7} \cos(\theta) v + \lr{\lambda_4 + \frac{1}{T_0^2} \pder{\tilde{\rho}}{\mu} \lambda_7} \cos^2(\theta) + \gamma_2 \sin^2(\theta) &= 0,\label{charge_2}\\
    6 \theta_1 \lr{\varepsilon_1 \varphi_5 - \varepsilon_5 \varphi_1} v^6 + \mathfrak{f}_1 \cos(\theta) v^5 + \mathfrak{f}_2 v^4 + \mathfrak{f}_3 \cos(\theta) v^3 + \mathfrak{f}_4 v^2 + \mathfrak{f}_5 \cos(\theta) v + \mathfrak{f}_6 &= 0,\label{long_caus}
\end{align}
\end{subequations}
where $\theta$ is the angle between $\kv$ and the $z$-axis, along which $\z_0^\m$ is aligned. The coefficients $\mf{f}_1,\ldots\mf{f}_6$ depend on $\theta$ in addition to the transport parameters and thermodynamic quantities. The first three equations in \eqref{caus_crit_1} are relatively easy to constrain such that $0 \leq v^2 < 1$. In particular, eq.~\eqref{shear_mode} is a shear mode which decouples due to the $SO(2)$ symmetry of the equilibrium state. The other two modes, eqs.~\eqref{charge_1} and \eqref{charge_2}, describe the large-$|\kv|$ behaviour of the perturbations to the transverse superfluid velocity $\delta \hat{\zeta}$ and the chemical potential $\delta \mu$ respectively.

In order to ensure that the roots of the quadratic $A x^2 + B x + C$ with $A > 0$  and $B, C \neq 0$ are such that $0 < x^2 < 1$ (for convenience, we exclude the possibility of non-propagating modes i.e.~$v$ vanishing identically, and purely luminal modes i.e.~$v$ equaling unity), the coefficients must obey the following properties (see e.g. \cite{Hoult:2020eho}):
\[
\Delta \equiv B^2-4AC > 0, \qquad B < 0, \qquad 0 < C < A, \qquad  A+B+C > 0.
\]
We can see immediately that for eqs. \eqref{shear_mode}, \eqref{charge_1}, and \eqref{charge_2}, these constraints cannot be satisfied for all $\theta$: we require $B < 0$, but they are all such that $B \propto \cos(\theta)$. Therefore, we must either have $B = 0$ or $C=0$. We choose to set the linear terms $B$ to zero in the first and third case via a choice of frame; for the second case, we set the constant term $C$ to zero (also by a choice of frame), as the thermodynamics does not guarantee the constant term in eq.~\eqref{charge_1} is of a definite sign for varying $\theta$. We therefore have
\begin{align}
&\text{Shear mode:}\left\{
\begin{aligned}
    &v^2 - \lr{\frac{\eta \sin^2(\theta)  - \varrho_3 \cos^2(\theta)}{\theta_1}} = 0,\\
    &\theta_3 + \varrho_1 = 0, \label{eq:shear_mode}\\
\end{aligned}\right.\\
&\text{Transverse superfluid velocity mode:}\left\{
\begin{aligned}
     &v\lr{v - \frac{\alpha_3}{\beta_3} \cos(\theta)} = 0,\\
    &\alpha_7 = 0,\label{eq:ssm}\\
\end{aligned}\right.\\
&\text{Charge mode:}\left\{
\begin{aligned}
    &v^2 + \frac{1}{\nu_2} \left[\lr{\lambda_4 + \frac{1}{T_0^2} \pder{\tilde{\rho}}{\mu} \lambda_7} \cos^2(\theta) + \gamma_2 \sin^2(\theta)\right] = 0,\\
    &\lambda_2 + \nu_4 + \frac{1}{T_0^2} \pder{\tilde{\rho}}{\mu} \nu_7 = 0.\label{eq:charge_mode}\\
\end{aligned}\right.
\end{align}
To enact the restrictions above, we impose the following equality-type constraints on the transport parameters that follow from eqs.~\eqref{eq:shear_mode}-\eqref{eq:charge_mode},
\be
\label{dec_causal_1}
\theta_3 = - \varrho_1, \quad \alpha_7 = 0, \quad \nu_7 = 0, \quad \lambda_2 = - \nu_4.
\ee
The inequality-type constraints that follow from eq.~\eqref{eq:shear_mode} are
\be
0 \leq \frac{\eta}{\theta_1} \leq 1, \qquad -1 \leq \frac{\varrho_3}{\theta_1} \leq 0\,,\label{eq:ineq_type_1}
\ee
the one following from eq.~\eqref{eq:ssm} is
\be
-1 \leq \frac{\alpha_3}{\beta_3} \leq 1\,,
\ee
while the ones from eq.~\eqref{eq:charge_mode} are
\be
-1 \leq \frac{\lambda_4 + \frac{1}{T_0^2} \pder{\tilde{\rho}}{\mu} \lambda_7}{\nu_2} \leq 0, \qquad -1 \leq \frac{\gamma_2}{\nu_2} \leq 0\,.\label{eq:ineq_type_2}
\ee
The subset of frames within the family of decoupled frames defined via eqs.\eqref{dec_causal_1}-\eqref{eq:ineq_type_2} therefore render six of the twelve modes in linearized superfluid hydrodynamics to be causal.

Let us now turn our attention to the order-6 polynomial in eq.~\eqref{long_caus}, which describes the large-$|\kv|$ behaviour of the two components of the fluid velocity in the $\kv$-$z$ plane and the temperature. Since the roots of an order-6 polynomial do not have closed-form analytic expressions in general, constraining the transport parameters in \eqref{long_caus} to ensure causality for arbitrary angle $\theta$ is rather difficult. In order to make an attempt at approaching this problem, we will consider three distinct cases, $\theta = 0$, $\theta = \frac{\pi}{2}$, and $\theta = \pi$, and ensure causality for all of them. In each of these cases, we will derive constraints to ensure that the roots of the ensuing polynomials imply causality. This will generically involve demanding that the polynomials be Schur-Cohn stable (i.e.~all roots lie within the open unit disk in the complex plane), and that all roots be real. In the case of the orthogonal limit $\theta = \frac{\pi}{2}$, we also enforce that the roots are positive, since the polynomial in this case is a cubic in $v^2$. For more details on how the constraints are derived for each case, refer to appendix \ref{app:Routh-Hurwitz}.

\subsubsection{$\theta = 0$: the co-aligned limit}
\label{sec:coaligned}
In general, the $SO(2)$ symmetry of the equilibrium state is broken by fixing $\kv$; hence, there is only one copy of the shear mode. In the case that $\theta = 0$, full $SO(2)$ rotational symmetry is restored in the setup, and a second copy of the shear mode \eqref{eq:shear_mode} (with $\theta = 0$) factors off of the order-6 polynomial eq.~\eqref{long_caus}. The remaining order-4 polynomial is easier to deal with, and has roots given by 
\begin{equation}\label{coaligned}
P_0(v) \equiv {\mathfrak a} v^4 + {\mathfrak b} v^3 + {\mathfrak c} v^2 + {\mathfrak d} v + {\mathfrak e} = 0
\end{equation}
where
\begin{align*}
    {\mathfrak a} &= 3 \lr{\varepsilon_5 \varphi_1 - \varepsilon_1 \varphi_5},\\
    {\mathfrak b} &= 6 \lr{\varepsilon_5 \pi_1 - \varepsilon_1 \pi_5} + 2 \lr{\varepsilon_1 \varphi_6 - \varepsilon_6 \varphi_1},\\
    {\mathfrak c} &= 4 \lr{\varepsilon_1 \pi_6 - \varepsilon_6 \pi_1} + 4 \lr{\varepsilon_1 \varphi_5 - \varepsilon_5 \varphi_1} + 6 \lr{ \pi_5 \varphi_1 - \pi_1 \varphi_5} + 2 \lr{\varepsilon_6 \varphi_5 - \varepsilon_5 \varphi_6},\\
    {\mathfrak d} &= 2 \lr{ \varepsilon_1 \pi_5 - \varepsilon_5 \pi_1} + 4 \lr{\varepsilon_6 \pi_5 - \varepsilon_5 \pi_6} + 4 \lr{\pi_1 \varphi_6 - \pi_6 \varphi_1} + 2 \lr{\varepsilon_6 \varphi_1 - \varepsilon_1 \varphi_6},\\
    {\mathfrak e} &= 2 \lr{\pi_1 \varphi_5 - \pi_5 \varphi_1} +4 \lr{ \pi_6 \varphi_5 - \pi_5 \varphi_6}+ \lr{\varepsilon_5 \varphi_1 - \varepsilon_1 \varphi_5} + 2 \lr{\varepsilon_5 \varphi_6 - \varepsilon_6 \varphi_5}.
\end{align*}
In order to derive constraints, we must demand that the roots of the quartic polynomial in eq.~\eqref{coaligned} be real, and within the open unit disk in the complex plane (we once again exclude the case of luminal propagation i.e.~$v = \pm 1$ for this analysis). This can be done, and yields the following set of constraints on the coefficients $\{\mathfrak{a},\mathfrak{b},\mathfrak{c},\mathfrak{d},\mathfrak{e}\}$:
\begin{align}
&\text{Real roots:}\left\{
\begin{aligned}
    &\Delta \equiv 256 s^3 - 128 q^2 s^2 + 144 q r^2 s + 16 q^4 s - 27 r^4 - 4 q^3 r^2 > 0,\\
    & \frac{q^2}{4} - s > 0, \label{eq:rr_1}\\
\end{aligned}\right.\\
&\text{Schur-Cohn stability:}\left\{
\begin{aligned}
    &\mc{S} \equiv \mf{a}+\mf{b}+\mf{c}+\mf{d}+\mf{e} >0,\\
    &2\mf{a} + \mf{b} - \mf{d} -2\mf{e} >0,\\
    &{\cal P} \equiv \lr{2 \mf{a} + \mf{b} - \mf{d} -2\mf{e}}\lr{6\mf{a}-2\mf{c}+6\mf{e}} - \mc{S} \lr{2\mf{a} -\mf{b} + \mf{d} -2 \mf{e}} > 0,\\
    &2\lr{2\mf{a} -\mf{b} +\mf{d} -2\mf{e}} {\cal P} - \lr{\mf{a} - \mf{b} + \mf{c} -\mf{d} +\mf{e}} \lr{6\mf{a} -2\mf{c}+6\mf{e}}^2 >0,\\
    &2\mf{a} - \mf{c} + 2\mf{e} > 0,\label{eq:schur_1}
\end{aligned}\right.
\end{align}
where
\begin{subequations}
    \begin{align}
        q &\equiv \frac{1}{\mf{a}^2} \lr{\mf{a} \mf{c} - \frac{3}{8} \mf{b}^2},\\
        r &\equiv\frac{1}{\mf{a}^3} \lr{\frac{\mf{b}^3 - 4 \mf{a} \mf{b} \mf{c}}{8} - \mf{d} \mf{a}^2},\\
        s &\equiv \frac{1}{\mf{a}^4} \lr{\mf{e} \mf{a}^3 - \frac{\mf{b}}{256} \lr{3 \mf{b}^3 + 16 \mf{abc} + 64 \mf{d} \mf{a}^2}},
    \end{align}
\end{subequations}
are the coefficients of the depressed quartic\footnote{For a generic polynomial of degree $n$, a change of variables can suppress the $(n-1)$-th order term. Such a polynomial is referred to as a ``depressed polynomial." In our case, this would correspond to the polynomial $v'^4 + q v'^2 + r v' + s = 0$, where the cubic term has been suppressed.} corresponding to eq.~\eqref{coaligned}. The conditions \eqref{eq:rr_1} and \eqref{eq:schur_1}, combined with eqs.~\eqref{dec_causal_1}-\eqref{eq:ineq_type_2}, provide us with a set of causal frames within the family of decoupled frames when $\theta = 0$.

\subsubsection{$\theta=\pi/2$: the orthogonal limit}
\label{sec:orthogonal}
In the case where $\theta = \f{\pi}{2}$, every odd-term in the order-6 equation \eqref{long_caus} goes to zero, since they are all proportional to $\cos(\theta)$. The order-6 polynomial in $v$ becomes an order-3 polynomial in $x \equiv v^2$, with the roots given by
\begin{equation}\label{orthogonal_limit}
P_{\pi/2}(x) \equiv {\mathfrak a} x^3 + {\mathfrak b} x^2 + {\mathfrak c} x + {\mathfrak d} = 0,
\end{equation}
where
\begin{align*}
    {\mathfrak a} &= 3 \theta_1 \lr{\varepsilon_5 \varphi_1 - \varepsilon_1 \varphi_5},\\
    {\mathfrak b} &= -\biggl[3 \varrho_1^2 \varepsilon_1 + \varepsilon_1 \lr{3 \theta_1 \varrho_3 - \lr{\theta_1 + 3 \eta} \varphi_5 + \lr{\pi_5 \varphi_6 - \pi_6 \varphi_5}} + \theta_1 \lr{\varepsilon_6 \varphi_5 - \varepsilon_5 \varphi_6 - 3 \pi_1 \varphi_5} + \pi_1 \lr{\varepsilon_6 \varphi_5 - \varepsilon_5 \varphi_6}\\
    &\quad+ \varphi_1 \lr{\varepsilon_5 \lr{3 \eta + \theta_1} + 3 \theta_1 \pi_5 + \varepsilon_5 \pi_6 - \varepsilon_6 \pi_5} + \varrho_1 \lr{3 \lr{ \varepsilon_5 \pi_1 - \varepsilon_1 \pi_5} + \lr{\varepsilon_6 \varphi_1 - \varepsilon_1 \varphi_6}}\biggr],\\
    {\mathfrak c} &= -\biggl[\varrho_1^2 \lr{\varepsilon_6 - 3 \pi_1 - \varepsilon_1} +  \varrho_3 \lr{\theta_1 \lr{\varepsilon_6 - 3 \pi_1 - \varepsilon_1} +\lr{\varepsilon_6 \pi_1 - \varepsilon_1 \pi_6} -  3\varepsilon_1 \eta} \\
    &\quad+ \varrho_1 \lr{\lr{\pi_1 \varphi_6 - \pi_6 \varphi_1} + \lr{\varepsilon_1 \pi_5 - \varepsilon_5 \pi_1} + \lr{\varepsilon_5 \pi_6 - \varepsilon_6 \pi_5} + 3 \eta \lr{\varepsilon_5 - \varphi_1}}\\
    &\quad+  \theta_1 \lr{ \lr{\pi_5 \varphi_6 - \pi_6 \varphi_5} + \lr{\pi_1 \varphi_5 - \pi_5 \varphi_1} - 3 \eta \varphi_5}\biggr],\\
    {\mathfrak d} &= \lr{3 \eta + \pi_6 - \pi_1} \lr{\varrho_1^2 + \theta_1 \varrho_3}.
\end{align*}
We need to constrain the roots of eq.~\eqref{orthogonal_limit} such that they are real, positive, and fall within the open unit disk. As such, we must have a positive discriminant, Schur-Cohn stability, and the Routh-Hurwitz stability of $-P_{\pi/2}(-x)$ (the overall $-$ is for convenience). This yields the following constraints:
\begin{align}
&\text{Real roots:}
\quad \Delta \equiv \mf{b}^2 \mf{c}^2 - 4 \mf{b}^3 \mf{d} + 18 \mf{abcd} - \mf{a} \lr{4 \mf{c}^3 + 27 \mf{ad}^2} >0, \label{eq:rr_2}\\
&\text{Schur-Cohn stability:}\left\{
\begin{aligned}
    &\mc{S}\equiv \mf{a}+\mf{b}+\mf{c}+\mf{d} >0,\\
    &3\mf{a} + \mf{b} -\mf{c} -3\mf{d} >0,\\
    &\mf{a}-\mf{b}+\mf{c}-\mf{d} >0,\\
    &\lr{3 \mf{a} + \mf{b} - \mf{c} - 3\mf{d}} \lr{3\mf{a} - \mf{b} -\mf{c} +3\mf{d}} - \mc{S} \lr{\mf{a}-\mf{b}+\mf{c}-\mf{d}} >0,\label{eq:schur_2} \\
\end{aligned}\right.\\
&\text{Positivity:}\quad \mf{a}>0, \quad \mf{b}< 0, \quad \mf{ad}-\mf{bc}>0, \quad \mf{d}<0.\label{positivity}
\end{align}
The criteria in eqs.~\eqref{eq:rr_2}-\eqref{positivity}, combined with eqs.~\eqref{dec_causal_1}-\eqref{eq:ineq_type_2}, yield a set of causal frames within the family of decoupled frames when $\theta = \frac{\pi}{2}$.

\subsubsection{$\theta=\pi$: the anti-aligned limit}
\label{sec:antialigned}
When $\theta = \pi$ and $\kv$ is anti-aligned with $\zeta_0^\m$, the $SO(2)$ rotational symmetry again gets restored. A $\theta = \pi$ copy of eq.~\eqref{eq:shear_mode} factors off, leaving an order-4 polynomial of the form
\[
{\mathfrak a} v^4 + {\mathfrak b} v^3 + {\mathfrak c} v^2 + {\mathfrak d} v + {\mathfrak e} = 0
\]
where
\begin{align*}
    {\mathfrak a} &= 3 \lr{\varepsilon_5 \varphi_1 - \varepsilon_1 \varphi_5},\\
    {\mathfrak b} &= 6 \lr{\varepsilon_1 \pi_5 - \varepsilon_5 \pi_1} + 2 \lr{\varepsilon_6 \varphi_1 - \varepsilon_1 \varphi_6},\\
    {\mathfrak c} &= 4 \lr{\varepsilon_1 \pi_6 - \varepsilon_6 \pi_1} + 4 \lr{\varepsilon_1 \varphi_5 - \varepsilon_5 \varphi_1} + 6 \lr{ \pi_5 \varphi_1 - \pi_1 \varphi_5} + 2 \lr{\varepsilon_6 \varphi_5 - \varepsilon_5 \varphi_6},\\
    {\mathfrak d} &= 2 \lr{ \varepsilon_5 \pi_1 - \varepsilon_1 \pi_5} + 4 \lr{\varepsilon_5 \pi_6 - \varepsilon_6 \pi_5} + 4 \lr{\pi_6 \varphi_1 - \pi_1 \varphi_6} + 2 \lr{\varepsilon_1 \varphi_6 - \varepsilon_6 \varphi_1},\\
    {\mathfrak e} &= 2 \lr{\pi_1 \varphi_5 - \pi_5 \varphi_1} +4 \lr{ \pi_6 \varphi_5 - \pi_5 \varphi_6}+ \lr{\varepsilon_5 \varphi_1 - \varepsilon_1 \varphi_5} + 2 \lr{\varepsilon_5 \varphi_6 - \varepsilon_6 \varphi_5}.
\end{align*}
Note the sign differences between this case and the co-aligned limit. The causality constraints are simply eqs.~\eqref{eq:rr_1} and \eqref{eq:schur_1}, but with different definitions of $\mathfrak{a}, \mathfrak{b}, \mathfrak{c}, \mathfrak{d}, \mathfrak{e}$.

\subsubsection{Summary of linearized causality constraints}
\begin{figure}[!ht]
    \centering
    \def\ImageScale{0.33}
    \includegraphics[scale=\ImageScale]{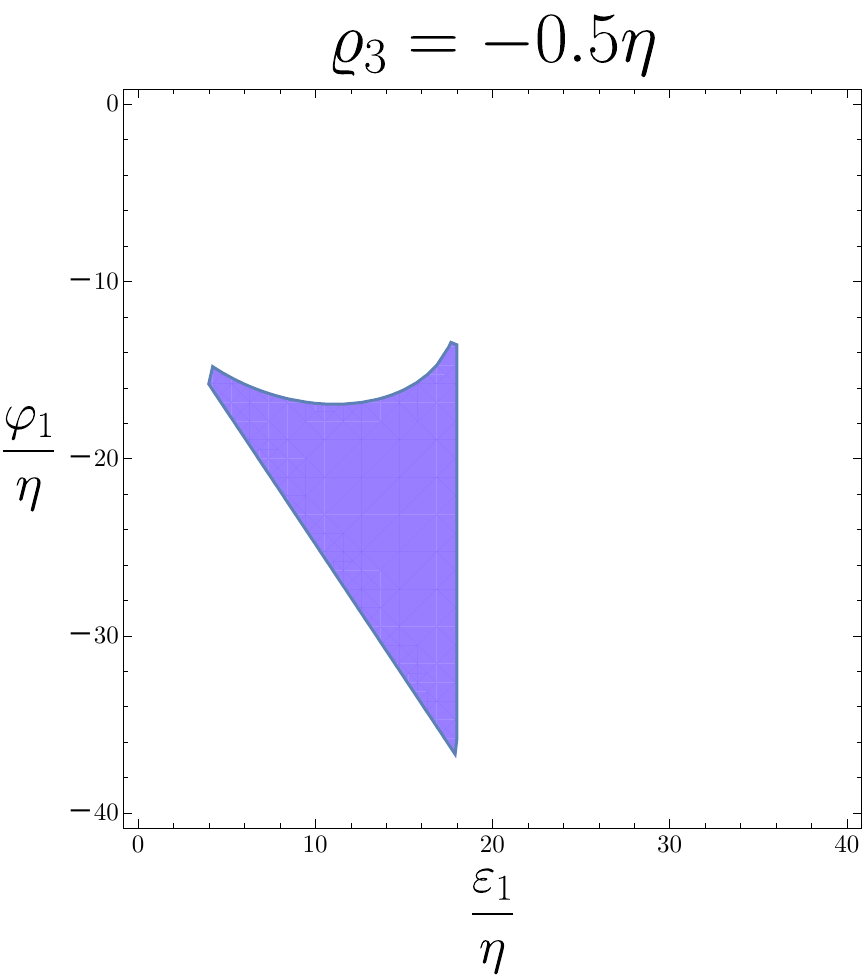}
    \includegraphics[scale=\ImageScale]{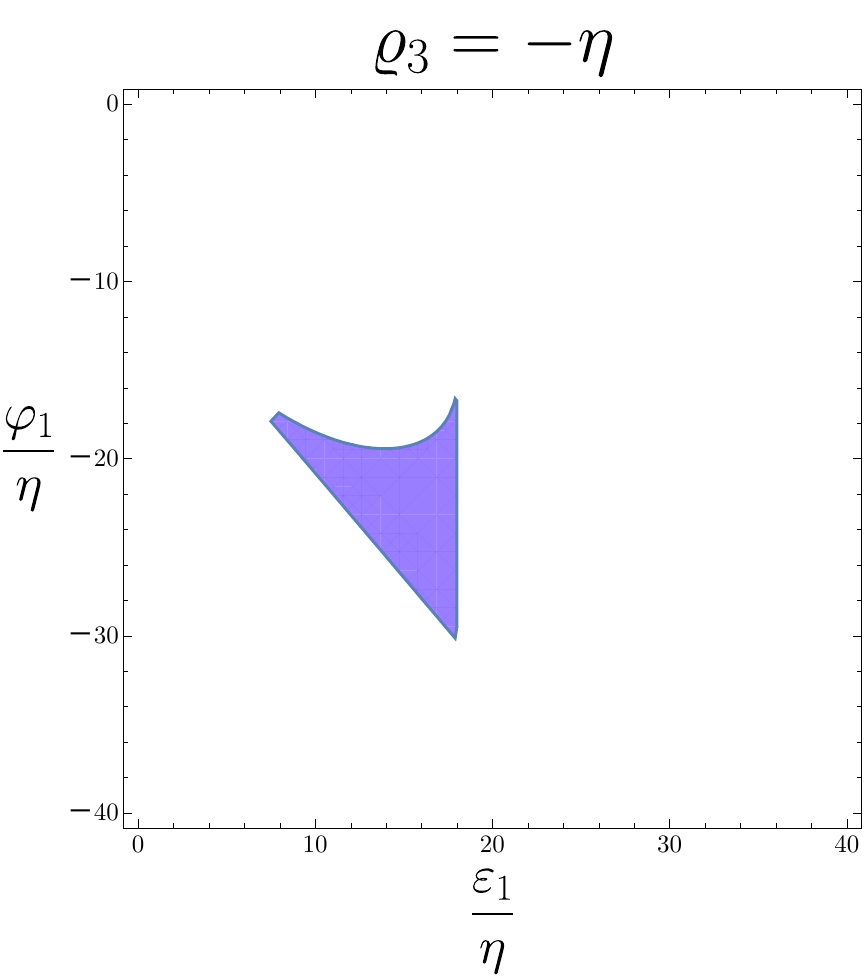}
    \includegraphics[scale=\ImageScale]{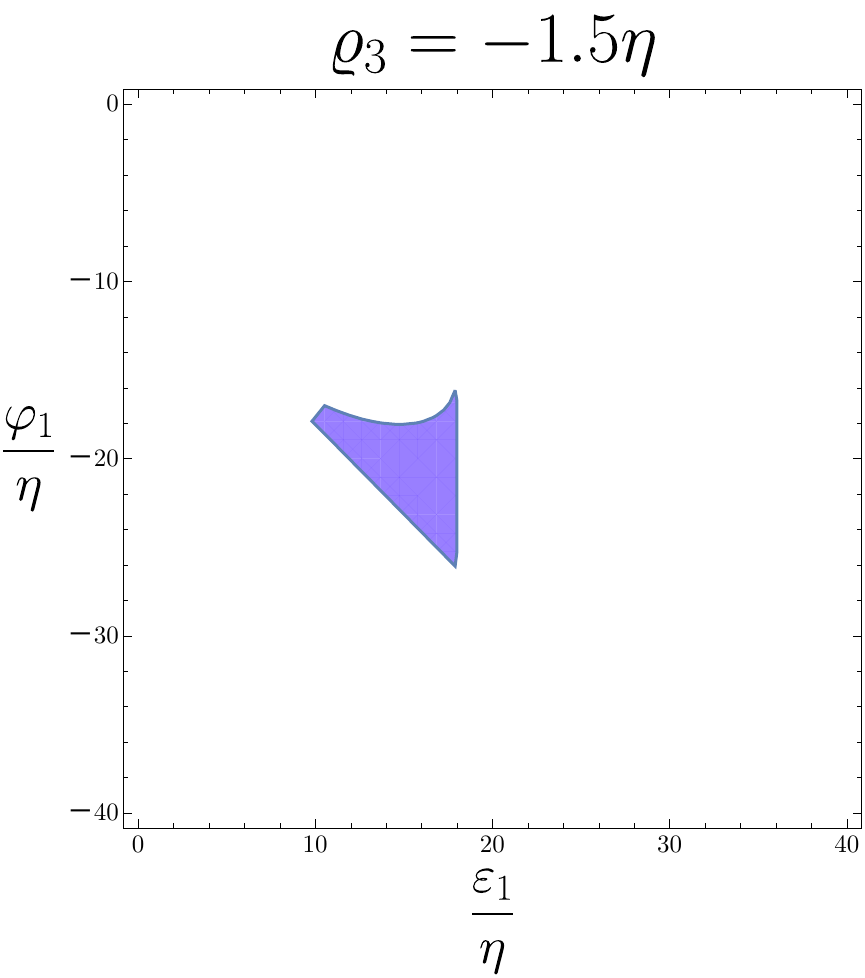}
    \includegraphics[scale=\ImageScale]{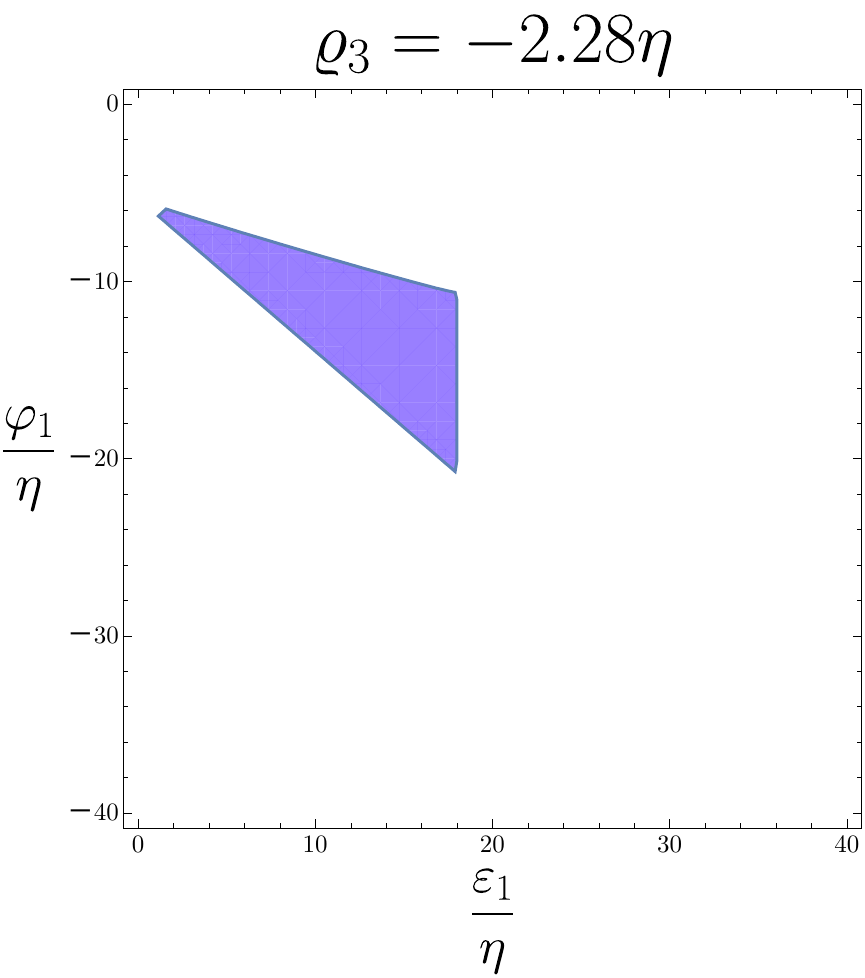}
    \includegraphics[scale=\ImageScale]{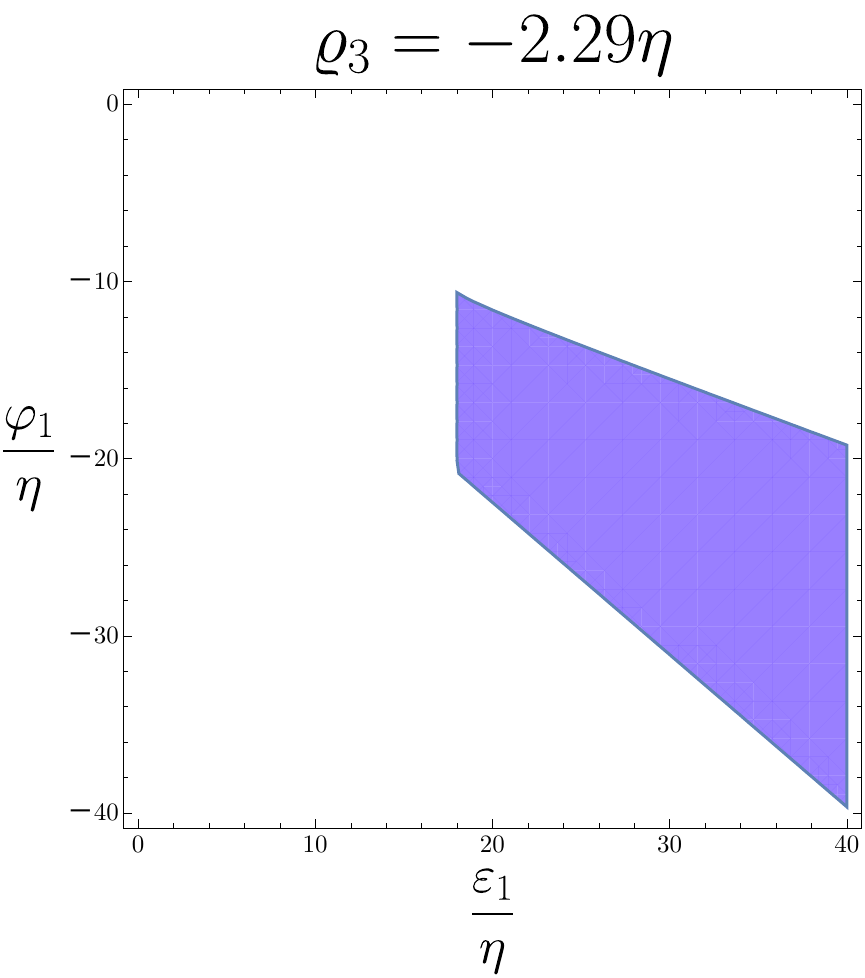}
    \includegraphics[scale=\ImageScale]{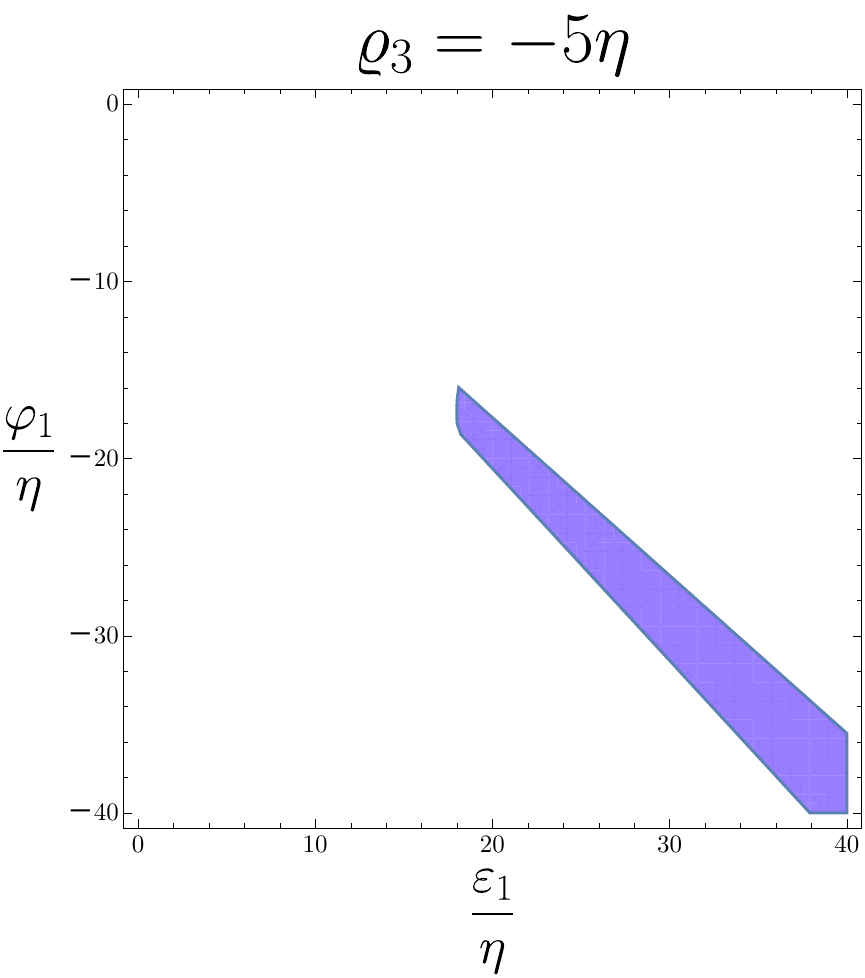}
    \caption{Regions in $(\ve_1/\eta, \vp_1/\eta)$ parameter space for different values of the ratio $M= (-\varrho_3)/\eta$ where the large-$|\kv|$ modes are causal for $\theta = 0$, $\pi/2$, and $\pi$ for the demonstrative choice eq.~\eqref{eq:plot_regions} within the extended decoupled family of frames.}
    \label{fig:causal_Tvec}
\end{figure}

Combining eqs.~\eqref{decoupled_frames} and \eqref{dec_causal_1}, let us refer to the following class of hydrodynamic frames as the ``Extended Decoupled Frame,"
\begin{equation}
\label{edf}
\begin{split}
    &\ve_{2,3,4,7} = \pi_{2,3,4,7} = \varphi_{2,3,4,7} = \theta_{2,4,5} = \varrho_{2,4,5} = 0, \quad \alpha_2 = \varsigma_2, \quad \alpha_4 = - \frac{1}{T_0^2}\pder{\tilde{\rho}}{\mu} \alpha_7,\\
    &\theta_3 = - \varrho_1, \quad \alpha_7 = 0, \quad \nu_7 = 0, \quad \lambda_2 = - \nu_4.
\end{split}
\end{equation}
Then, from eqs.~\eqref{eq:ineq_type_1}-\eqref{eq:ineq_type_2}, the shear mode, the charge mode, and the transverse superfluid velocity mode are all causal if the following constraints are obeyed:
\begin{equation}
\label{lin_caus_simp}
\begin{split}
    &0 \leq \frac{\eta}{\theta_1} \leq 1, \quad -1 \leq \frac{\varrho_3}{\theta_1} \leq 0,\\
    &0 \leq \frac{\lambda_4 + \frac{1}{T_0^2} \pder{\tilde{\rho}}{\mu} \lambda_7}{\nu_2} \leq 1, \quad 0 \leq \frac{\gamma_2}{\nu_2} \leq 1,\\
    &-1 \leq \frac{\alpha_3}{\beta_3} \leq 1.
\end{split}
\end{equation}
The temperature-velocity mode eq.~\eqref{long_caus} is rendered causal for the set $\theta = \{0, \pi/2, \pi\}$ when a complicated set of constraints are satisfied, as discussed in secs.~\ref{sec:coaligned}-\ref{sec:antialigned}. In particular, the constraints depend on $\eta$ in a non-trivial way, which is entirely out of our hands. However, we can state the constraints in terms of $\eta$ and $\varrho_3$, and the ratio between them: $M = (-\varrho_3)/\eta \geq 0$. For instance, in the following demonstrative frame,
\begin{align}
\label{eq:plot_regions}
\ve_5 &= 0,\quad \pi_5 = 0,\quad \varrho_5 = 0, \quad \ve_6 = \ve_1/2,\quad \pi_6 =  \ve_1/6,\quad \varphi_6 = \varphi_1/2,\quad \varrho_6 = 0,\quad \pi_1 = \ve_1/3, \nonumber\\
\alpha_3 &= \beta_3/2, \quad \gamma_2 = - \nu_2/2,  \quad \lambda_4 = - \lr{\pder{\tilde{\rho}}{\mu} \lambda_7/T_0^2 +\nu_2/2},\quad \varphi_5 = - \pi \eta,\nonumber\\
\theta_1 &= 2 \lr{\eta - \varrho_3},\quad 
\varrho_1 = -\frac{\sqrt{1+2M}}{2}\lr{\eta - \varrho_3},
\end{align}
the causality of the mode for the three cases $\theta = \{0, \pi/2, \pi\}$ is dependent on the value of $M$. For visualization, we have plotted the region of causality for six different choices of the ratio $M$ in figure \ref{fig:causal_Tvec} for the ratios of the transport parameters $\ve_1/\eta$ and $\vp_1/\eta$. Provided eq.~\eqref{lin_caus_simp} holds, each point in the shaded regions in these plots corresponds to a frame where linear causality is respected for $\theta = \{0, \pi/2, \pi\}$.

\section{Discussion and outlook}
\label{sec:discussion}
In this paper, we have extended the BDNK formalism to first order relativistic conformal superfluid hydrodynamics. The novel feature that comes when considering the superfluid state are the antisymmetric equations \eqref{eq_for_xi}, some of which are dynamical while the rest act as constraints. The oft used LLCP frame for superfluid hydrodynamics has unstable and acausal modes. To cure this, we recast superfluid hydrodynamics in a general frame, and demanded that perturbations about the equilibrium state be linearly stable and causal. Linear stability in the $|\kv|\to0$ limit could be satisfied (restricting to the \emph{decoupled frames} reported in eq.~\eqref{decoupled_frames}) if the transport parameters satisfied the requirements in eq.~\eqref{lin_stab_2}. This is a representative choice, and is by no means exhaustive. To further respect linear causality, we introduced a subset of frames, titled the \emph{extended decoupled frames}, eq.~\eqref{edf}, where a subset of the causality criteria on the transport parameters take a particularly simple form, eq.~\eqref{lin_caus_simp}. However, the complimentary subset of the causality criteria depends sensitively on the choice of the angle $\theta$ between the wave vector $\kv$ and the direction of the spacelike vector $\z^\m_0$. We chose three different values of $\theta$ and worked out the criteria that need to be met for causality to hold. A more general analysis, valid for arbitrary $\theta$, will likely require a fully numerical approach, which is beyond the scope of the present work.  

Several comments are in order. Firstly, even though we focused on conformal superfluids, the formalism should be extendable to non-conformal superfluids in a straightforward manner. Secondly, the linear stability criteria we have worked out ensure stability of the gapped modes as $|\kv|\to 0$. Ensuring linear stability for arbitrary $\kv$, though tedious, can be done with knowledge about the thermodynamic properties of the system (such as the equation of state) as well as resorting to a numerical approach. The same could be said about ensuring causality for arbitrary angle $\theta$.

Let us now consider potential future directions. One fruitful approach to pursue would be to build upon the ideas presented in \cite{Hoult:2021gnb}, which discusses how the hydrodynamic frame ambiguity arises in a kinetic theory-based approach, applicable when an underlying quasiparticle description is available. To arrive at a causal theory, a vital role is then played by the zero modes of the collision operator in the Boltzmann equation. Further, \cite{Hoult:2021gnb} elucidates how a similar story plays out via the zero modes of the bulk differential operator in the holographic fluid-gravity correspondence \cite{Bhattacharyya:2007vjd}. Exploring how these ideas play out for relativistic superfluid hydrodynamics would be an interesting and insightful exercise.

It would also be worthwhile to pursue the line of thought presented in \cite{Bhattacharyya:2024ohn, Bhattacharyya:2024jxm}, where the authors argue that by a suitable all-orders frame transformation one can arrive at a hydrodynamic frame where the spectrum of linearized perturbations about the equilibrium state consists entirely of the physical hydrodynamic modes, eliminating the non-hydrodynamic modes completely. Generalizing their construction to relativistic superfluids would be an interesting exercise.

Further, the superfluids we have discussed in the present work are on a flat background; it would be very interesting to extend the discussion to superfluids coupled to dynamical gravity. The cores of neutron stars are posited to be superfluidic in nature~\cite{MIGDAL1959655,BAYM1969}, and therefore a general relativistic causal theory seems relevant. Secondly, a more detailed description of superfluids should also include the effects of vortices.

Additionally, we have assumed that the superfluid under investigation is in the regime far removed from the critical point, and the dynamics of the ``amplitude mode'' \cite{Pekker-Varma, Donos:2022xfd} can be neglected. It would be interesting to see how the inclusion of the amplitude mode into our analysis, relevant when the system is near criticality, will affect the causality and stability aspects. 

Finally, in this paper we performed a linearized analysis of the equations of motion for a relativistic superfluid. An important outstanding question is the classification of the non-linear system of partial differential equations. Are the BDNK superfluid equations strongly hyperbolic? Do they admit a well-posed initial value problem? What is the role of the constraint equations present in \eqref{eq_for_xi}? Following the discussion in~\cite{Hoult:2024qph}, it is likely that a linearized analysis is sufficient to determine non-linear causality of the equations of motion; however, in the present paper we explicitly solved the constraint equations. The effect this has on the equivalence between linear and non-linear causality described in~\cite{Hoult:2024qph} remains to be seen. The causality of relativistic magnetohydrodynamics \cite{Armas:2022wvb, Hoult:2024qph} is likely a useful guide, as it is also a system containing constraint equations (namely, $\nabla{\cdot}B=0$).

The questions described in this discussion are research directions to which we plan to return. In the meantime, we hope this investigation into stable and causal viscous superfluid hydrodynamics will spur further interest in the properties of these systems.

\acknowledgments
We would like to thank Pavel Kovtun for several helpful discussions and thoughtful suggestions that helped better our understanding and improved the presentation of this manuscript. The work of REH was supported in part by the NSERC of Canada. This work was supported in part by the National Science Foundation under Grant No. NSF PHY-1748958. The work of AS was supported in part by the European Research Council (ERC) under the European Union’s Horizon 2020 research and innovation programme (grant agreement no.\ 758759). REH would like to acknowledge the hospitality received from the University of Illinois at Urbana-Champaign, Oxford University, and \'Ecole Polytechnique (CPHT), who hosted him while he completed parts of this work. AS would like to acknowledge the hospitality received from the Korea Institute for Advanced Study, Seoul, and Nordita, Stockholm, where parts of this work were done. 

\appendix
\section{Further details about the LLCP frame}
\label{further_LLCP}
\subsection{Positivity of entropy production}
\label{entropy_LLCP}
In this appendix, we perform an entropy current analysis ensuring positive semi-definiteness of entropy production onshell in the LLCP frame for a conformal superfluid, section \ref{LLCPframe}, which will in turn impose constraints on the LLCP frame transport coefficients. We start by computing the divergence of the vector $su^\m$, which is given by
\be\label{eq:divsbasic}
\begin{split}
\del_\m \left(s u^\m\right) &= u^\m \del_\m s + s\, \del\cdot u\\
&= u^\m \left(\f{1}{T} \del_\m \e - \f{\m}{T} \del_\m \rho + \f{\tr}{T} \del_\m \z\right) + \f{1}{T} \left(\e + P - \m \rho\right) \del\cdot u,
\end{split}
\ee
where we have made use of the thermodynamic relations eqs.\ \eqref{Gibbs-Duhem} and \eqref{etedefs}. Next, we need to go onshell and make use of the hydrodynamic equations to simplify the expression above. Recall that the LLCP frame constitutive relations for a conformal superfluid have the form
\begin{subequations}
\label{LLCP_const_app}
\begin{align}
T^{\m\n} &= \e u^\m u^\n + \left(\e-(d-1)P\right) z^\m z^\n + P \tilde{\D}^{\m\n} + 2 \m \tilde{\rho} u^{(\m} z^{\n)} + \Pi^{\m\n}\, ,\label{LLCP_stress}\\
J^\m &= \rho u^\m + \tilde{\rho} z^\m \, ,\label{LLCP_curr}\\
\xi^\m &= - \left(\mu+\mc{M}_1\right) u^\m + \z z^\m\, ,
\label{LLCP_xi}
\end{align}
\end{subequations}
where the derivative corrections $\Pi^{\m\n}$ and $\mc{M}_1$ have the form given by eq.\ \eqref{full_set_LLCP} in terms of our chosen basis of independent $\mc{O}(\del)$ Weyl covariant scalars, vectors and tensors. More concretely,
\begin{subequations}
\label{LLCP_Diss_Terms}
\begin{align}
\Pi^{\m\n} &= (\D^{\m\n} -d z^\m z^\n)  \sum_{n=5}^7 \bar{\pi}_n \mbb{s}_n + 2 u^{(\m} z^{\n)} \sum_{n=5}^7 \bar{\vp}_n \mbb{s}_n + 2 \sum_{n=1,3} \bar{\th}_n \mbs{v}_n^{(\m} u^{\n)} \nonumber\\
&\quad + 2\sum_{n=1,3} \bar{\varrho}_n \mbs{v}_n^{(\m} z^{\n)} - \eta \s^{\m\n}_u\, ,\label{LLCP_Diss_Term1}\\
\mc{M}_1 &= \sum_{n=5}^7 \bar{\a}_n \mbb{s}_n\, . \label{LLCP_Diss_Term2}
\end{align}
\end{subequations}
Utilizing the constitutive relations eq.\ \eqref{LLCP_const_app}, one gets the hydrodynamic equations 
\begin{subequations}
	\begin{align}
		u_\n \del_\m T^{\m\n} = 0 &\Rightarrow u^\m\del_\m \e = - (\e+P) \del\cdot u +\z\tilde{\rho} \, z^\m z^\n \del_\m u_\n - \tilde{\rho}\, z^\l \del_\l \m \nonumber\\
		&\qquad\qquad\quad\,- \m\,  z^\l \del_\l \tilde{\rho}  + \m \tilde{\rho} \left( u^\m u^\n \del_\m z_\n - \del\cdot z\right) + u_\n \del_\m \Pi^{\m\n},\\
		\del_\m J^\m = 0 &\Rightarrow u^\m \del_\m \r = - \r \, \del\cdot u - \tilde{\r}\, \del\cdot z - z^\m \del_\m \tilde{\r} \, ,\\
		u^\m z^\n \del_{[\m} \xi_{\n]} = 0 &\Rightarrow u^\m \del_\m \z = - (\m + \mc{M}_1) u^\m u^\n \del_\m z_\n + z^\n \del_\n (\m + \mc{M}_1) - \z z^\m z^\n \del_\m u_\n\, .
	\end{align}
\end{subequations}
Making use of these equations to eliminate the quantities in the first parentheses of eq.\ \eqref{eq:divsbasic}, we get
\be\label{eq:suggestive}
\del_\m(su^\m) = \f{1}{T} u_\n \del_\m \Pi^{\m\n} - \f{\tilde{\rho}}{T} \, \mc{M}_1\, u^\m u^\n \del_\m z_\n + \f{\tilde{\rho}}{T} \, z^\l \del_\l \mc{M}_1.
\ee 
Eq.\ \eqref{eq:suggestive} suggests that we can define the entropy current $J^\m_S$ as
\be\label{eq:ent_current}
J^\m_S \equiv s u^\m - \f{1}{T} u_\n \Pi^{\m\n} - \f{\tilde{\rho}}{T}\, \mc{M}_1 \, z^\m.   
\ee
By virtue of eq.\ \eqref{eq:suggestive}, the divergence of the entropy current $J^\m_S$ is
\be\label{eq:divScurrent}
\del_\m J^\m_S = - \Pi^{\m\n} \del_\m\left(\f{u_\n}{T}\right) - \mc{M}_1 \, \D^{\m\n} \del_\m \left(\f{\tilde{\rho} \, z_\n}{T}\right).
\ee
The expression above expresses the divergence of the entropy current in terms of the dissipative structures that enter the constitutive relations in the LLCP frame. For physical configurations, the divergence of the entropy current should be positive semidefinite.
Putting in the expressions for $\Pi^{\m\n}$ and $\mc{M}_1$ from eq.\ \eqref{LLCP_Diss_Terms} into eq.\ \eqref{eq:divScurrent}, we get
\be
\label{eq:dmuJmuS}
\begin{split}
\del_\m J^\m_S = &- \frac{1}{T} \bar{\vp}_5 \mbb{s}_5^2 + \frac{d}{T} \bar{\pi}_6 \mbb{s}_6^2 - T^{d-1} \bar{\a}_7 \mbb{s}_7^2 + \frac{1}{T} \left(d\bar{\pi}_5 - \bar{\vp}_6\right) \mbb{s}_5 \mbb{s}_6 - \frac{1}{T} \left(\bar{\vp}_7 + T^d \bar{\a}_5\right) \mbb{s}_5 \mbb{s}_7 \\
&+\frac{1}{T} \left(d\bar{\pi}_7 - T^d \bar{\a}_6\right) \mbb{s}_6 \mbb{s}_7 - \frac{\bar{\th}_1}{T} \, \mbs{v}_1 \cdot \mbs{v}_1 - \frac{1}{T} \left(\bar{\th}_3 + \bar{\varrho}_1\right) \mbs{v}_1 \cdot \mbs{v}_3 - \frac{\bar{\varrho}_3}{T}\, \mbs{v}_3 \cdot \mbs{v}_3 \\
&+ \frac{\eta}{2T} \, \s^{\m\n}_u \s_{u\, \m\n}\, .
\end{split}
\ee
We next impose the Onsager reciprocity condition,%
\footnote{
The Onsager relations are the condition that retarded two-point Green's functions are covariant under time-reversal, see e.g.~\cite{Kovtun:2012rj}.
}
which leads to the demand that the coefficients of the cross terms between independent scalars and vectors in the expression above be identical. This immediately leads to the relations given in eq.\ \eqref{Onsager} of the main text, leading to a reduction in the number of independent transport parameters from 14 to 10. The remaining 10 parameters form the set of physical transport coefficients for the conformal superfluid at $\mc{O}(\del)$ in the derivative expansion. These are further constrained by the requirement of positive semidefiniteness of entropy production i.e., $\del_\m J^\m_S \ge 0$. Mathematically, this translates to ensuring that the RHS of eq.\ \eqref{eq:dmuJmuS} is a positive semidefinite quadratic form. This in turn leads to the constraints quoted in eq.\ \eqref{2ndLawConstraints} on the LLCP frame transport coefficients.

\subsection{Relations between physical transport coefficients \& transport parameters}
\label{LLCP_Rels}
We know that for a conformal superfluid there are 10 physical transport coefficients. How, then, are these 10 physical transport coefficients related to the 70 transport parameters that appear in eq.~\eqref{full_set}? Here, we broadly sketch the relevant equations.

To begin with, in the main body of the text, several frame invariant quantities were defined. Let us decompose the frame-invariants eqs.~\eqref{eq:frame-invar-vectors},~\eqref{eq:frame-invar-scalars} with respect to the basis of scalars and vectors eqs.~\eqref{scalar_data_conf},~\eqref{vector_data_conf}:
\begin{equation}
    \begin{gathered}
       \textbf{Scalars:}\quad {\cal G} = \sum_{n=1}^7 g_n \mbb{s}_n, \quad {\cal L} = \sum_{n=1}^7 \ell_n \mbb{s}_n, \quad {\cal H} = \sum_{n=1}^7 h_n \mbb{s}_n\,,\\
       \textbf{Vectors:}\quad {\cal L}^\mu = \sum_{n=1}^5 L_n \mbs{v}^\mu, \quad {\cal K}^\mu = \sum_{n=1}^5 k_n \mbs{v}^\mu\,.
    \end{gathered}
\end{equation}
Then, using eq.~\eqref{full_set}, we have
\begin{subequations}\label{eq:frame-invar-trans-param}
    \begin{align}
        g_n &= \pi_n - \mbs{b}_1 \varepsilon_n - \mbs{b}_2 \nu_n - \mbs{b}_3 \lambda_n - \mbs{b}_4 \beta_n\,,\\
        \ell_n &= \varphi_n - \mbs{c}_1 \varepsilon_n - \mbs{c}_2 \nu_n - \mbs{c}_3 \lambda_n - \mbs{c}_4 \beta_n\,,\\
        h_n &= \alpha_n - \mbs{d}_1 \varepsilon_n - \mbs{d}_2 \nu_n - \mbs{d}_3 \lambda_n - \mbs{d}_4 \beta_n\,,\\
        L_n &= \theta_n - \mbs{e}_1 \gamma_n - \mbs{e}_2 \varsigma_n\,,\\
        k_n &= \varrho_n +\tilde{\rho} \varsigma_n\,.
    \end{align}
\end{subequations}
Suppose we now work in the LLCP frame, eq.~\eqref{LLCP_choice}. We therefore see that
\begin{equation}
\begin{gathered}\label{eq:ap_LLCP_no_ideal}
    {\cal P}_1 = \sum_{n=1}^7 g_n \mbb{s}_n, \quad {\cal U}_1 = \sum_{n=1}^7 \ell_n \mbb{s}_n, \quad {\cal M}_1 = \sum_{n=1}^7 h_n \mbb{s}_n\,,\\
    {\cal Q}^\mu_1 = \sum_{n=1}^5 L_n \mbs{v}_n^\mu, \quad {\cal R}_1^\mu = \sum_{n=1}^5 k_n \mbs{v}_n^\mu, \quad {\cal T}_1^{\mu\nu} = - \eta \sigma^{\mu\nu}_u\,.
\end{gathered}
\end{equation}
One may now apply the ideal-order equations of motion, \eqref{eq:scalar_rels} and \eqref{eq:vector_rels}, writing $\mbb{s}_{1,2,3,4}$ in terms of $\mbb{s}_{5,6,7}$, and $\mbs{v}^\mu_{2,4,5}$ in terms of $\mbs{v}_1^\mu$ (recall that $\mbs{v}_3^\mu$ does not enter the ideal-order equations). Let us schematically write
\begin{equation}
    \begin{gathered}\label{eq:ideal_eq_app}
        \mbb{s}_n = a_n \mbb{s}_5 + b_n \mbb{s}_6 + c_n \mbb{s}_7 \quad \lr{n \in [1,2,3,4]}\,,\\
        \mbs{v}_n^\mu = d_n \mbs{v}_1^\mu \quad \lr{n \in [2,4,5]}\,,
    \end{gathered}
\end{equation}
where $a_n,\,b_n,\,c_n,\,d_n$ are thermodynamic functions. Then, applying the solutions~\eqref{eq:ideal_eq_app} to the decompositions~\eqref{eq:ap_LLCP_no_ideal}, we can write
\begin{subequations}
\begin{align}
    {\cal P}_1 &=  \lr{g_5 +\sum_{n=1}^4 a_n g_n}\mbb{s}_5 + \lr{g_6 +\sum_{n=1}^4 b_n g_n}\mbb{s}_6 + \lr{g_7 +\sum_{n=1}^4 c_n g_n}\mbb{s}_7\,,\\
    {\cal U}_1 &=  \lr{\ell_5 +\sum_{n=1}^4 a_n \ell_n}\mbb{s}_5 + \lr{\ell_6 +\sum_{n=1}^4 b_n \ell_n}\mbb{s}_6 + \lr{\ell_7 +\sum_{n=1}^4 c_n \ell_n}\mbb{s}_7\,,\\
    {\cal M}_1 &=  \lr{h_5 +\sum_{n=1}^4 a_n h_n}\mbb{s}_5 + \lr{h_6 +\sum_{n=1}^4 b_n h_n}\mbb{s}_6 + \lr{h_7 +\sum_{n=1}^4 c_n h_n}\mbb{s}_7\,,\\
    {\cal Q}_1^\mu &=  \lr{L_1 + d_2 L_2 + d_4 L_4 + d_5 L_5} \mbs{v}_1^\mu + L_3 \mbs{v}_3^\mu\,,\\
    {\cal R}_1^\mu &=  \lr{k_1 + d_2 k_2 + d_4 k_4 + d_5 k_5} \mbs{v}_1^\mu + k_3 \mbs{v}_3^\mu\,.
\end{align}
\end{subequations}
By direct comparison to the LLCP frame transport coefficients in eq.~\eqref{LLCPframe}, then, we can state that
\begin{subequations}\label{eq:LLCP_trans_con_frame_invar}
    \begin{align}
        \bar{\pi}_5 &= g_5 +\sum_{n=1}^4 a_n g_n, \quad \bar{\pi}_6 = g_6 +\sum_{n=1}^4 b_n g_n, \quad \bar{\pi}_7 = g_7 +\sum_{n=1}^4 c_n g_n\,,\\
        \bar{\varphi}_5 &= \ell_5 +\sum_{n=1}^4 a_n \ell_n, \quad \bar{\varphi}_6 = \ell_6 +\sum_{n=1}^4 b_n \ell_n, \quad \bar{\varphi}_7 = \ell_7 +\sum_{n=1}^4 c_n \ell_n\,,\\
        \bar{\alpha}_5 &= h_5 +\sum_{n=1}^4 a_n h_n,\quad \bar{\alpha}_6 = h_6 +\sum_{n=1}^4 b_n h_n, \quad \bar{\alpha}_7 = h_7 +\sum_{n=1}^4 c_n h_n\,,\\
        \bar{\theta}_1 &= L_1 + d_2 L_2 + d_4 L_4 + d_5 L_5,\quad \bar{\theta}_3 = L_3\,,\\
        \bar{\varrho}_1 &= k_1 + d_2 k_2 + d_4 k_4 + d_5 k_5,\quad \bar{\varrho}_3  = k_3\,.
    \end{align}
\end{subequations}
The LLCP frame transport coefficients are then given in terms of the general-frame transport parameters by substituting equations~\eqref{eq:frame-invar-trans-param} into equations~\eqref{eq:LLCP_trans_con_frame_invar}. The Onsager reciprocity relations eq.~\eqref{Onsager} will then provide further relations between these quantities.

\section{Routh-Hurwitz criterion, Schur-Cohn stability, and their applications}
\label{app:Routh-Hurwitz}
In section \ref{sec:new_material}, we made use of the Routh-Hurwitz (RH) criterion and Schur-Cohn stability. In this appendix, we give their mathematical description for completeness.

\subsection{The Routh-Hurwitz criterion}
Let us begin by considering a polynomial of arbitrary order $n$, whose roots we are interested in:
\begin{equation}\label{ap:RH_poly}
    P(x) \equiv a_n x^n + a_{n-1} x^{n-1} + ... + a_1 x + a_0 = 0.
\end{equation}
For our purpose, we assume that all the coefficients $a_n$ are real. This is the case in the analysis of causality, but not stability, due to the anisotropy of the equilibrium state (which allows for odd powers of $|\kv|$ in the spectral curve). The real case was derived by Routh \cite{routh1877treatise} and Hurwitz \cite{Hurwitz1895}; see \cite{korn2013mathematical} for more details. An extension for the case of complex coefficients is presented in \cite{Hastir_2023}.

Let us split the polynomial $P(x)$ into two sub-polynomials: one containing even powers, and one containing odd powers,
\begin{subequations}
    \begin{align}
        P_0 = a_n x^n + a_{n-2} x^{n-2} + ...\, ,\\
        P_1 = a_{n-1} x^{n-1} + a_{n-3} x^{n-3} + ...\, .
    \end{align}
\end{subequations}
We now define a new polynomial $P_2$, which is defined to be the remainder after dividing $P_0$ by $P_1$,
\begin{equation}
    P_2 \equiv {\rm Rem}\lr{P_0, P_1}.
\end{equation}
From this polynomial, we can define a new polynomial $P_3$, defined by
\begin{equation}
    P_3 \equiv {\rm Rem}\lr{P_1, P_2}.
\end{equation}
One can continue in this fashion until one reaches a zeroth-order polynomial; at that point, the procedure terminates. For a polynomial with non-zero coefficients, this will occur at $P_n$. Let us denote the highest-order coefficient of a polynomial by ${\cal P}(P)$. Then the Routh-Hurwitz criterion amounts to demanding that in the set 
\begin{equation}
\{{\cal P}(P_0), {\cal P}(P_1), ..., {\cal P}(P_n)\}
\end{equation}
all elements have the same sign. For an example of this process applied to a number of polynomials, the interested reader may refer to Appendix B of \cite{Hoult2020:thesis}.

\subsection{Schur-Cohn Stability}
Schur-Cohn stability demands that the roots of a polynomial lie within an open disc of unit radius centred at the origin~\cite{Zahreddine1992}. It is simply given by demanding the Routh-Hurwitz stability of the polynomial \eqref{ap:RH_poly} after performing a M\"obius transformation,
\begin{equation}
    P'(x) = (x-1)^n P\lr{\frac{x+1}{x-1}} = 0.
\end{equation}
This transformation maps the unit disc to the left-half of the complex plane, allowing us to straightforwardly apply the RH-criterion discussed in the previous section. This is how the causality constraints of section \ref{gen_causality} were derived. The demand that the disk be open amounts to demanding that no roots lie on the imaginary axis after the transformation; in particular, we demand that $P(1) \neq 0$.

\section{Acausality of the Modified Landau Frame}
\label{mod_Landau_frame}
The LLCP frame discussed in section \ref{LLCPframe} was shown to be acausal, and therefore unstable (see section \ref{LLCP_problems}). However, there are other important superfluid hydrodynamic frames as well. An example is the so-called ``Landau frame" of~\cite{Arean:2023nnn}. To avoid confusion with the LLCP frame, as well as the usual Landau frame used in relativistic hydrodynamics, we will refer to this as the \emph{modified Landau frame}. This frame is defined by the choices
\begin{equation}
{\cal E}_1 = {\cal U}_1 = {\cal N}_1 = {\cal Z}_1 = 0, \quad {\cal Q}_1^{\mu} = {\cal X}_1^\mu = 0\,. 
\end{equation}
This is equivalent to 
\begin{equation}
u_\mu T_1^{\mu\nu} = 0, \quad u_\mu J_1^\mu = 0, \quad \xi_\mu  = -\lr{\mu + {\cal M}_1} u_\mu + \zeta_\mu\,.
\end{equation}
The first-order corrections to the stress-energy tensor, charge current, and phase field are then given by
\begin{subequations}
\begin{align}
T^{\mu\nu}_1 &= - \lr{d-1} {\cal P}_1 z^\mu z^\nu + {\cal P}_1 \tilde{\Delta}^{\mu\nu} + 2 {\cal R}_1^{(\mu} z^{\nu)} + {\cal T}_1^{\mu\nu}\, ,\\
J^\mu_1 &=  {\cal S}_1 z^\mu + {\cal J}^\mu_1\, ,\\
\xi^\mu &= -{\cal M}_1 u^\mu\, .
\end{align}
\end{subequations}
After demanding positivity of entropy production and imposing the Onsager relations, one finds that the non-vanishing dissipative corrections are given by
\begin{subequations}
    \begin{align}
        {\cal P}_1&= - \frac{\tilde{\lambda}_6 T}{d} \mbb{s}_4 + \tilde{\pi}_6 \mbb{s}_6 - \frac{T^d \tilde{\alpha}_6}{d} \mbb{s}_7\,,\\
        {\cal R}_1^\mu &= T \tilde{\gamma}_3 \mbs{v}_2^\mu + \tilde{\varrho}_3 \mbs{v}_3^\mu\,,\\
        {\cal T}_1^{\mu\nu} &= - \tilde{\eta} \sigma^{\mu\nu}\,,\\
        {\cal S}_1 &= \tilde{\lambda}_4 \mbb{s}_4 + \tilde{\lambda}_6 \mbb{s}_6 + T^{d-1} \alpha_4 \mbb{s}_7\,,\\
        {\cal J}^\mu_1 &= \tilde{\gamma}_2 \mbs{v}_2^\mu + \tilde{\gamma}_3  \mbs{v}_3^\mu\,,\\
        {\cal M}_1 &= \tilde{\alpha}_4 \mbb{s}_4 + \tilde{\alpha}_6 \mbb{s}_6 + \tilde{\alpha}_7 \mbb{s}_7.
     \end{align}
\end{subequations}
Performing a linearized analysis in this frame, which falls into the class of ``constraint frames," eq.~\eqref{eq:constraint_frames}, we find that there are six hydrodynamic modes, and zero non-hydrodynamic modes, which clearly violates the third condition of eq.~\eqref{large_k_caus_constraints}. Indeed, we find that of the six hydrodynamics modes, five behave as $\omega \sim k^2$ in the limit of large $k$, violating the first two conditions of eq.~\eqref{large_k_caus_constraints}. The sixth goes as $\omega \sim z{\cdot}k$. This mode is therefore acausal, and there exists a boosted reference frame in which it will exhibit instability. This instability, unlike the physical instability described in~\cite{Arean:2023nnn}, is observer dependent, and not physical.

\bibliographystyle{JHEP}
\bibliography{superfluid-refs}

\end{document}